\documentclass[10pt]{article}

\usepackage[margin=1in]{geometry}

\usepackage{amsmath}

\usepackage{moreverb}
\usepackage[usenames,dvipsnames]{xcolor}

\usepackage{hyperref}

\usepackage[inline]{enumitem}

\usepackage{graphicx}
\usepackage{caption}
\usepackage{subcaption}

\usepackage{algorithm}
\usepackage{algpseudocode}
\algnewcommand\algorithmicforeach{\textbf{for each}}
\algdef{S}[FOR]{ForEach}[1]{\algorithmicforeach\ #1\ \algorithmicdo}

\usepackage{booktabs}
\newcommand{\ra}[1]{\renewcommand{\arraystretch}{#1}}

\newcommand{\best}[1] {\textbf{#1}}
\newcommand{\worst}[1] {\textit{#1}}

\title{Database Engines: Evolution of Greenness}
\author{
    Andriy~V.~Miranskyy \\
    Department of Computer Science \\
    Ryerson University \\
    Toronto, ON, M5B 2K3,  Canada \\
    avm@ryerson.ca
    \and
    Zainab~Al-zanbouri \\
    Department of Computer Science \\
    Ryerson University \\
    Toronto, ON, M5B 2K3,  Canada \\
    \and
    David Godwin \\
    IBM Toronto Software Lab, \\
    8200 Warden Ave., \\
    Markham, ON, L6G 1C7, Canada
    \and
    Ayse~Basar~Bener \\
    Department of Mechanical and Industrial Engineering \\
    Ryerson University \\
    Toronto, ON, M5B 2K3, Canada \\    
}

\begin{document}

\maketitle

\begin{abstract}

\textbf{Context:} Information Technology consumes up to 10\% of the world's electricity generation, contributing to CO$_2$ emissions and high energy costs. Data centers, particularly databases, use up to 23\% of this energy. Therefore, building an energy-efficient (green) database engine could reduce energy consumption and CO$_2$ emissions.

\textbf{Goal:} To understand the factors driving databases' energy consumption and execution time throughout their evolution.

\textbf{Method:} We conducted an empirical case study of energy consumption by two MySQL database engines, InnoDB and MyISAM, across 40 releases. We examined the relationships of four software metrics to energy consumption and execution time to determine which metrics reflect the greenness and performance of a database.

\textbf{Results:} Our analysis shows that database engines' energy consumption and execution time increase as databases evolve. Moreover, the Lines of Code metric is correlated moderately to strongly with energy consumption and execution time in 88\% of cases.

\textbf{Conclusions:} Our findings provide insights to both practitioners and researchers. Database administrators may use them to select a fast, green release of the MySQL database engine. MySQL database-engine developers may use the software metric to assess products' greenness and performance. Researchers may use our findings to further develop new hypotheses or build models to predict greenness and performance of databases. 

\end{abstract}


\maketitle

\section{Introduction}\label{sec:intro}

Information Technology (IT) energy requirements are significant. Literature shows that 1500 Terawatt hours
\footnote{Terawatt hour $= 10^{12}$ watt hours.}
(TWh) per year (or 10\% of the worldwide energy generation) are consumed by IT \cite{mills2013cloud}. With the adoption of the Cloud paradigm, data centers consume up to 23\% (350 TWh \cite{mills2013cloud}) of the overall amount of energy that is used by IT. In the US alone, it is expected that the data center energy consumption will grow from 91 TWh in 2013 to 139 TWh in 2020, with the energy bill rising from \$9.0 billion USD to \$13.7 billion USD \cite{whitney2014data}. 

Not only energy consumption is expensive, but it also affects the environment. The average monthly concentration of CO$_2$ in atmosphere reached 400 parts per million in 2015 -- the highest in the past 800,000 years~\cite{jackson2015}. It is estimated that world energy-related CO$_2$ emissions will increase from 32.3 billion metric tons in 2012 to 35.6 billion metric tons in 2020 and to 43.2 billion metric tons in 2040~\cite{eia_tr2016}. Data centers in US emitted 97 million metric tons of CO$_2$ in 2013, it is expected that the emissions will grow to 147 million metric tons by 2020 \cite{whitney2014data}. Research in environmental sustainability focuses on economic, environmental, social, and human drivers that impact the environment and human beings. In this respect, IT, in general, and software, in particular, may contribute to environmental sustainability through the development of environmentally friendly systems. This may happen in different ways, e.g., using energy efficiently via decreasing the used resources, resulting, in turn, in the reduced CO$_2$ emissions. Moreover, IT processes can be made more sustainable by decreasing the energy consumption and the negative emissions of companies and individuals. 
Green IT is focused on studying these issues. Formally, Green IT is a study and practice of efficient use of computing resources to decrease the negative impact on the environment \cite{murugesan2008harnessing}, it can be applied to various high-tech domains, such as data centers, mobile computing and embedded systems  \cite{kern2011green}. However, in this paper, we will focus on the data center domain.

Obviously, the energy in the data centers is consumed by hardware. However, since hardware is driven by software, software is responsible for consuming energy as well \cite{kern2011green}. The software side has received less attention than the hardware side and a few solutions (such as energy test suites \cite{noureddine2012preliminary, wilke2013jouleunit, tpc_energy}) have been put forth to comprehend energy efficiencies. 

Software manufacturers are now paying more attention to making the enterprise software more energy efficient (greener). This is driven not only by external requests from their clients (that need to reduce maintenance cost of their data centers), but also by requests from within the software manufacturing organization \cite{Brown:2010:TEC:1666420.1666438}. The majority of software manufacturers (e.g., Amazon, Google, IBM, Microsoft, Oracle, and SAP) offer their products via Cloud using Platform as a Service (PaaS) or Software as a Service (SaaS) delivery models. Internal technology operations teams, managing the PaaS and SaaS offerings, are also adding the voice to the chorus of the clients, asking to reduce energy expenses for internal data centers. 

Software development teams may benefit from making their own software greener too. Adoption of continuous delivery and integration practices, such as nightly builds and automatic unit and regression tests, based on literature~\cite{Duarte2006, Winter:2015} and authors' experience, requires hundreds of build and test servers per product. This computing power is required to build and run tens of thousands of test suites on multiple platforms to support multiple releases of a given product (maintained by the development team). Thus, by making their own software greener, developers can decrease their expenses, by reducing energy bill for the internal server farms.

Databases are considered to be the backbone in the software world; hence, they are responsible for a significant proportion of the overall software energy consumption. Therefore, we focus on understanding how to make databases more efficient (greener) by trying to identify the main factors that affect database energy consumption and execution time. 
In addition, a lot of database engines (especially relational ones) are mature products that have been developed for decades (e.g., first Oracle database was release in 1978, IBM DB2 -- in 1983, and MySQL -- in 1995). Multiple releases of these products are available and maintained by their development teams. Thus, it is important to understand how database engine greenness evolves from release to release.

\textbf{Research objective:} We aim to identify how the energy consumption and execution time of a database engine may change when the database evolves from one release to another, in addition to understanding how these changes are related to some database-associated properties such as raw data size, database engine type (MyISAM/InnoDB), and database memory size.
Furthermore, this research investigates the software metrics that can have direct effects on the sustainability development in a database represented by its effect on energy consumption and execution time. By using such metrics, software developers can assess greenness of their products by measuring characteristics of their code, thus eliminating the need to execute a reference workload against a software product and to measure energy consumption directly (in turn, saving time and resources). In order to achieve our research objective, we focus on answering two \textbf{research questions}:
\begin{description}
\item[RQ1] How does the energy consumption and execution time of a database engine change as the product matures (from one release to another)?
\item[RQ2] Which software metrics reflect energy consumption and execution time?
\end{description}

Answering RQ1 will help us identify the factors leading to green and, hopefully, fast databases. Answering RQ2 will help us build models that can predict software greenness and performance based on software metrics that can be easily extracted from source code (such as code size or code churn metrics). This information should be of interest to practitioners, since software vendors such as Apple, IBM and Microsoft are seeking more sustainable products with lower levels of energy consumption and  execution time \cite{hindle2012green, Hindle2015, Brown:2010:TEC:1666420.1666438}. It should also be of interest to researchers, since the information can be used to build universal models of software's energy consumption and performance.  

In order to answer these research questions, in this paper, we study energy consumption and execution time across 40 different releases (shipped between 2005\footnote{The earliest available in MySQL archive.} and 2014) of two database engines (MyISAM and InnoDB) of the MySQL database. MySQL is the most commonly used and most popular open source database in the world \cite{mysql}. We chose MySQL as the software under study, because the MySQL database is a mature product (having been developed since 1995) with a large (approximately 1 million lines of code) codebase being actively developed. This gives us enough data to study the product's evolution.

To answer RQ1, we study the relation between energy consumption (or execution time) for all the MySQL versions\footnote{In this paper, we use the terms `version' and `release' interchangeably.} under study. To answer RQ2, we examine the relation between software metrics from one side and energy consumption (or execution time) 
from the other side. 

This type of work required building a framework to automate all the necessary processes such as database installation, upgrade from version to version, executing the workload, reading and collecting the measurements from the special measurement meter and recording the measurements for all the MySQL releases used in addition to creating a database for all the experimental data results. It also required building a system to extract software metrics from the code base of MySQL so that the relation between the metrics on one side and energy consumption (or execution time) on the other side could be established. All these requirements are addressed in this work.

This is the first study examining the relation between different MySQL database releases and their energy consumption as well as execution times. This research differs from the previous research \cite{hindle2012green, Hindle2015} that has examined the link between different versions of Firefox web browser and their performance. In addition, to the best of our knowledge, this work is the first study to establish a link between MySQL databases' energy consumption and their execution time from one side and the software metrics (namely Lines of Code (LOC), Lines of Code Changed (LOCC), and Traditional/Modified Cyclomatic Complexity (TCC/MCC)), from another side.

The rest of the paper is structured as follows.  Related work is discussed in Section~\ref{sec:background}. Methodology of our experiments is explained in Section~\ref{sec:methodology}. Section~\ref{sec:results} provides the results of our experiments. Threats to validity are given in Section~\ref{sec:threats}. Finally, Section~\ref{sec:conclusions} concludes the paper.

\section{Related Work}\label{sec:background}
This section is structured as follows. Energy-related research, focused on computer parts and operating system (OS)-level software, is provided  in Section~\ref{sec:lit:hw_sw}. Energy-related benchmarks and frameworks are discussed in Section~\ref{sec:lit:bench}. Relevant database-related research is given in Section~~\ref{sec:lib:db}. Finally, research related to mining software repositories and energy consumption data is shown in Section~\ref{sec:lib:msr}.

\subsection{Energy consumption: Hardware and OS-level Software}\label{sec:lit:hw_sw}

A number of researchers have focused on energy consumption in IT. Delaluz et al. \cite{delaluz2001dram} conducted a comprehensive study of software and hardware systems to determine the benefit of the DRAM mode control abilities for energy savings. They addressed an essential issue in energy saving for mobile and computing environments by specifically concentrating on the memory system, which consumes around 90\% of the complete energy consumed by the system when ignoring input/output processes \cite{delaluz2001dram}. 

Tiwari et al. \cite{tiwari1996instruction} presented the power usage of a single CPU. They defined an assessment-based instruction-level power analysis method, which provides an accurate and practical way of measuring the power cost of software and describes an assessment-based instruction-level power analysis method that makes it possible to effectively analyze software power consumption. Mittal et al. \cite{mittal2012empowering} presented an energy simulation tool that allows developers to estimate the energy use for their mobile apps on their development workstation itself. 

There are several studies about the power consumption of devices. Bircher et al. \cite{bircher2007complete} produced power models for the complete system depending on processor performance events. Greenwalt et al. \cite{greenawalt1994modeling}  measured and modeled the power consumption of hard drives. The hard disk state model provides both the quantitative data and insight necessary to design an efficient power management system. Stemm et al. \cite{stemm1997measuring} studied two types of optimization (namely, transport-level and application-level) of network interfaces to decrease their energy consumption.

Li et al. \cite{li1994quantitative} performed a quantitative analysis of the costs and benefits of spinning down a disk drive as a power management technique. The main idea behind the power consumption measurement movement is to be followed by suggestions or actions taken in order to find solutions to any undesirable outcomes. Selby et al. \cite{selby2011unconventional} applied methods to analyze the relationship between global variable usage and the efforts required by software maintenance and examined the effects of optimizations upon power usage. Fei et al. \cite{fei2007energy} employed source code change techniques to decrease the energy overheads accompanying application/OS connections and modified the source code changes and compiler optimizations in order to reduce power usage. Feng et al. \cite{feng2005power} introduced a framework for studying the power-performance efficiency of the NAS parallel benchmarks on a 32-node Beowulf cluster. 

\subsection{Benchmarks and Frameworks}\label{sec:lit:bench}
Some researchers have concentrated on the idea of benchmarking and examining power measurement. Asmel et al. \cite{amsel2010green} described a tool that approximates the energy consumption of software in order to help concerned consumers make knowledgeable decisions about the software they use. Gurumurthi et al. \cite{gurumurthi2002using} introduced a complete system power simulator that represents the CPU, the hierarchy of memory and a low-power disk subsystem and calculates the power performance of both side applications and the OS. 

Researchers have also developed frameworks for measuring and testing energy consumption. For example, Noureddine et al. \cite{noureddine2012preliminary} built runtime energy monitoring framework, enabling easy reporting on the energy consumption of system processes. Wilke et al. \cite{wilke2013jouleunit} created a generic framework for software energy profiling and testing.

\subsection{Database-related research}\label{sec:lib:db}

Researchers have studied changes to the design of a database engine, but not to its energy consumption. For example, Shang et al. \cite{shang2014exploratory} investigated changes to the amount of communicated information passed to system administrators over multiple versions of the PostgreSQL database engine and the Hadoop data processing framework.

Researchers designed prototypes of energy-aware database management systems. Chen et al. \cite{chen2013green} designed ReinDB database engine that, in the presence of renewable and non-renewable energy sources, distributes database workload to minimize usage of the non-renewable energy source. Liu et al. \cite{liu2013generating} created optimizer for execution plans of queries sent to the database engine; the optimizer minimizes energy consumption for a given query. In addition to these prototypes, Transaction Processing Performance Council created guidelines for measuring energy consumption of database workloads \cite{tpc_energy}.

\subsection{Mining Software Repository and Energy Consumption}\label{sec:lib:msr}

The closest to our work are \cite{hindle2012green, Hindle2015, gupta2011energy}. Gupta et al. \cite{gupta2011energy} study focused on combining Mining Software Repository (MSR) \cite{kagdi2007survey} techniques with power performance and presented the first study from a software engineering perspective on energy awareness problems. The authors of \cite{gupta2011energy} introduced a method for gathering and analyzing power data on mobile devices running Windows Phone 7. Their methodology describes and quantifies power consumption, detects differences in power consumption and predicts power consumption. The work by \cite{gupta2011energy} is complementary to ours, because it focuses on examining the power consumption in different modules (a module is a part of a program) within the same software (Windows Phone 7) and finding which module consumes the most power. Moreover, it focused on finding the typical energy shape patterns of certain modules. We, on the other hand, are focusing on multiple versions of the same product (MySQL). Additionally, we concentrate on understanding the relation between energy consumption (or execution time) and the product development of MySQL.

Hindle \cite{hindle2012green, Hindle2015} demonstrated combining the MSR research and energy consumption by studying multiple versions of the Firefox web browser regarding characteristic energy consumption patterns of multiple modules of the web browser. He also examined the relation between the LOC and LOCC software metrics and energy consumption. These works are complementary to ours, because we focus on a different product (database instead of web browser) and study the effect of multiple software metrics (LOCC, MCC and TCC, in addition to LOC) on energy consumption and execution time. 

The works cited above demonstrate the significance and importance of the study of power consumed by software in various areas of IT.

\section{ Methodology and Experiments }\label{sec:methodology}
In constructing the experiment we followed guidelines of Wohlin et al.~\cite{wohlin2012} in general and Hindle~\cite{hindle2012green, Hindle2015} in particular, with minor variations. Design of the experiments capturing energy used, time spent, and system's statistics is given in Section~\ref{sec:framework}; extraction of software metrics -- in Section~\ref{sec:softm}; and analysis of the data -- in Section~\ref{sec:analysis}.

\subsection{ Experimental Design }\label{sec:framework}

Product selection, as well as selection of a set of versions under study, is described in Section~\ref{sec:sus}; testbed setup is given in Section~\ref{sec:testbed}; our test case is discussed in Section~\ref{sec:workload}; product configuration is shown in Section~\ref{sec:product_cfg}; automation of test is depicted in Section~\ref{sec:auto}; details of instrumentation and measurements are provided in Section~\ref{sec:instr}; and effect of baseline energy consumption on the test case measurement -- in Section~\ref{sec:baseline}.

\subsubsection{Software Under Study}\label{sec:sus}

\begin{figure}[ht]
    \centering
    \includegraphics[scale = 0.8]{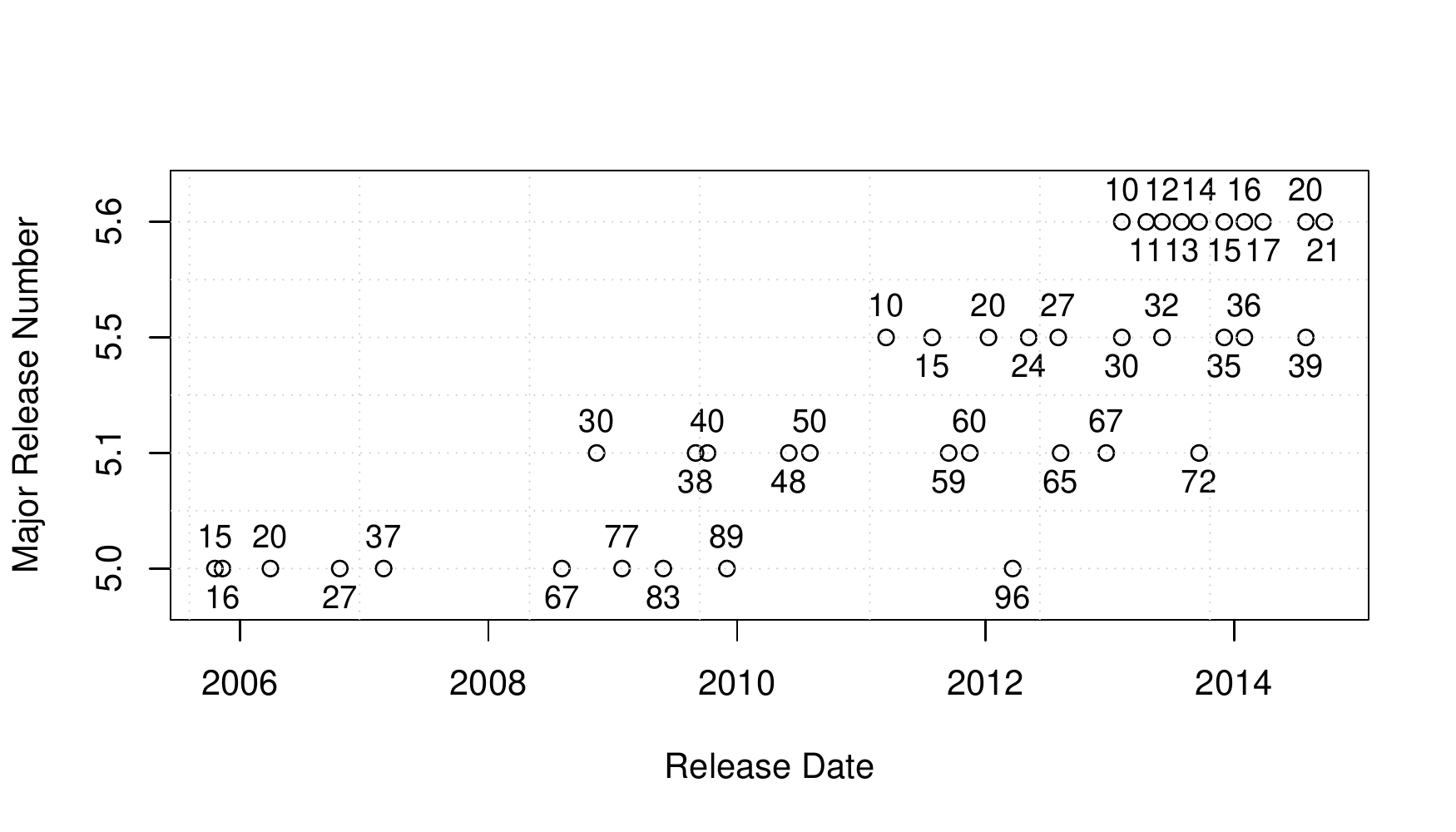}
    \caption{Time line of the releases. $Y$-axis depicts major release numbers; values above the points -- minor release numbers. For example, number 21 in the top right corner represents release date for version 5.6.21.
}
    \label{fig:timeline}
\end{figure}

As discussed in Section~\ref{sec:intro}, our software under study are 40 different releases (shipped between 2005 and 2014) of two database engines (MyISAM and InnoDB) of the MySQL database. Table~\ref{tab:releases} shows the list of the MySQL major releases used in our experiments along with a list of the minor MySQL releases under study. Shipping timeline of the releases is given in Figure~\ref{fig:timeline}.

We chose MySQL as the software under study, because the MySQL database is a mature and popular product~\cite{mysql} (having been developed since 1995) with a large (approximately 1 million lines of code) codebase being actively developed. These facts make MySQL a good candidate to study product evolution.

\begin{table}[ht]
    \centering
    \begin{tabular}{@{}lcccc@{}}\toprule
     Major release                           & v.5.0    & v.5.1    & v.5.5    & v.5.6    \\
     \midrule
    ~                                        & v.5.0.15 & v.5.1.30 & v.5.5.10 & v.5.6.10 \\
    ~                                        & v.5.0.16 & v.5.1.38 & v.5.5.15 & v.5.6.11 \\
    ~                                        & v.5.0.20 & v.5.1.40 & v.5.5.20 & v.5.6.12 \\
    ~                                        & v.5.0.27 & v.5.1.48 & v.5.5.24 & v.5.6.13 \\
    Minor releases                           & v.5.0.37 & v.5.1.50 & v.5.5.27 & v.5.6.14 \\
    ~                                        & v.5.0.67 & v.5.1.59 & v.5.5.30 & v.5.6.15 \\
    ~                                        & v.5.0.77 & v.5.1.60 & v.5.5.32 & v.5.6.16 \\
    ~                                        & v.5.0.83 & v.5.1.65 & v.5.5.35 & v.5.6.17 \\
    ~                                        & v.5.0.89 & v.5.1.67 & v.5.5.36 & v.5.6.20 \\
    ~                                        & v.5.0.96 & v.5.1.72 & v.5.5.39 & v.5.6.21 \\
    \bottomrule
    \end{tabular}
    \caption{A list of the four major MySQL releases under study with their corresponding ten minor versions used in the experiments.}
    \label{tab:releases}
\end{table}

\subsubsection{Testbed setup}\label{sec:testbed}

The computer used in our experiments has two Intel Pentium 4 HT 630 3GHz CPUs, 3GB of RAM, 320GB of storage on a magnetic hard drive Western Digital WD3200AAKS. In order to eliminate effect of changing environmental conditions, this machine was rack-mounted in a data centre of the Department of Computer Science, Ryerson University. The data centre is thermostated at $29^{\circ}$C (with 40\% humidity) by industrial-grade air conditioners.

The operating system installed on the machine was Ubuntu Linux OS v.14.04 with v.3.13.0-32-generic x86\_64 kernel, Server edition. We chose the Linux platform because it is a de facto standard server platform, and it is better designed for capturing computer-related statistics. Moreover, the server edition (unlike desktop edition) has smaller number of programs preinstalled~\cite{ubuntu_server_vs_desktop}. For example, our installation did not have graphical user interface. This leads to a smaller number of programs that may run in the background and affect the results. The computer was dedicated to our workloads -- no other tasks\footnote{Excluding limited number of crucial daemons ran by the OS; e.g., \texttt{logrotate} daemon performs archival and rotation of OS logs on a daily basis.}  were executed on this machine concurrently. We discuss effect of the daemons in Section~\ref{sec:baseline}. No monitor was attached to the machine; all communications with the machine happened remotely via ssh.

\subsubsection{Test scenario}\label{sec:workload}
\begin{figure}[t]
    \centering
    \includegraphics[scale = 0.6]{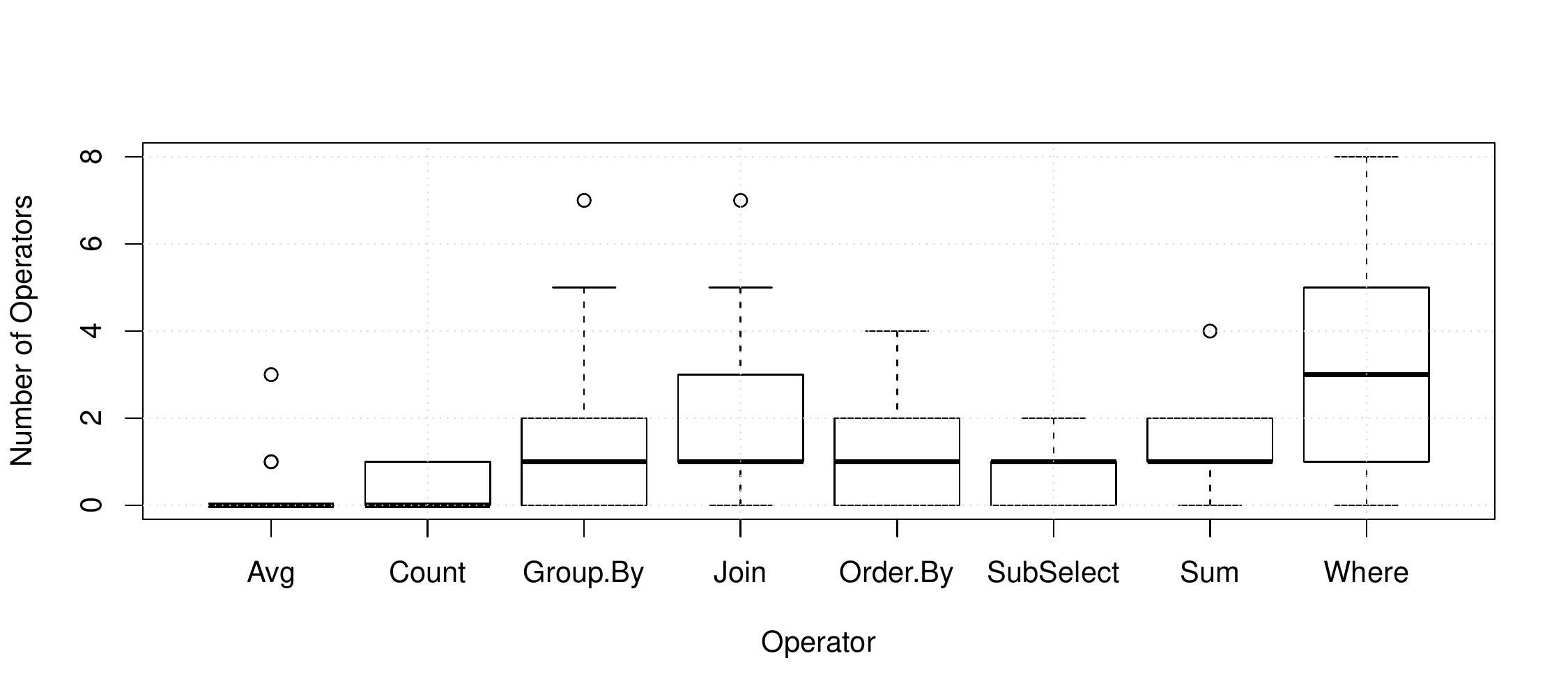}
    \caption{Distribution of operators and/or parameters in the queries of the TPC-H workload. One query has 35 `Where' operators; however, we truncate $y$-axis at 8 to improve readability.}
    \label{fig:query_distr}
\end{figure}

As a reference database workload, we used Transaction Processing Performance Council Benchmark H (TPC-H) version 2.17.0 \cite{tpc_h}. It is considered as the standard benchmark for analytic workloads in the database community \cite{tpc_h}. This benchmark has a set of 22 business-oriented ad-hoc queries. The data and queries simulate business practices and requirements. This benchmark mimics the decision support systems that use large volume of data, execute complicated queries (with relatively low volume of transactions), and answer business-related questions. We provide distribution of the number of operators in the queries in Figure~\ref{fig:query_distr}. In-depth technical analysis of the queries is given in \cite{boncz2014}.

\subsubsection{Product configuration}\label{sec:product_cfg}

In our eight experiments, we loaded 1GB and 3GB of raw data into the database. Given the size of our hardware platform, 1GB dataset is used to mimic in-memory processing workload (since all the data can be stored in memory and no I/O operations have to be performed once the data are loaded into memory); the 3GB workload mimics workload operating on a large dataset (that cannot fit into memory and require intense I/O operations). The raw data is generated by DBGEN tool from the TPC-H package \cite{tpc_h}. 

Different buffer pool sizes: Tuning the MySQL default installation is very important to improve its performance, and the key buffer cache is an essential element to be changed in this tuning process; a key buffer is used to cache the data from the hard drive into the memory \cite{mysql_innodb, mysql_myisam}. Theoretically, the more memory that is allocated to the cache, the faster the data processing. We set the buffer cache value to either 256MB or 1024MB. 

Table~\ref{tab:experiments} shows the eight different experimental setups with the corresponding memory buffer size; numbers in brackets represent the raw data size used in each experiment.

Other configuration parameters: in general, there exists infinite number of combinations of parameters. We followed best practices, as suggested by the official documentation (e.g., see~\cite{mysql_best_practice}).

\begin{table}[t]
    \centering
    \begin{tabular}{@{}ccc@{}} \toprule
     Memory buffer size & MyISAM             & InnoDB               \\
     \midrule
    256MB               & Experiment 1 (1GB) & Experiment 5 (1GB)   \\
    ~                   & Experiment 2 (3GB) & Experiment 6 (3GB)   \\
    \addlinespace
    1024MB (1GB)        & Experiment 3 (1GB) & Experiment 7 (1GB)   \\
                        & Experiment 4 (3GB) & Experiment 8 (3GB)   \\
    \bottomrule
    \end{tabular}
    \caption{A list of the experiments with the corresponding memory buffer size; numbers in brackets represent the amount of raw data used in each experiment.}
    \label{tab:experiments}
\end{table}

\subsubsection{Test Automation}\label{sec:auto}

We designed a framework that automates experimentation process following, conceptually,  the guidelines of Hindle~\cite{hindle2012green, Hindle2015}. In particular, our framework automates the installation and measurement process: it installs the specific MySQL version (obtained from \cite{mysql}), upgrades and configures (as discussed in Section~\ref{sec:product_cfg}) the database, and executes the 22 reference TPC-H queries \footnote{We used TPC-H utility called QGEN to generate these queries.} one by one. The framework measures the execution time, system state statistics (such as CPU and I/O load), and energy consumption. We provide pseudo code of the framework in Appendix~\ref{sec:experiment_details}\footnote{Note that each experiment is repeated three times, increasing accuracy and precision, and reducing measurement error.}. Details of the measurement process are given below.

\subsubsection{Instrumentation and Measurements}\label{sec:instr}

The device we used to measure energy consumption is called ``Watts up? PRO"~\cite{watts_up}. The device allows direct reading of the measurements to a computer. Energy measuring device can measure both power and energy. We chose to measure energy (with resolution of 0.1 Wh), as per recommendations of the support personnel of the manufacturer of the device (obtained by e-mail). Based on the recommendation and the manual~\cite{wattsup_manual}, the accuracy of the power (watts) measurement is $\pm 1.5\%$ (partially attributed to the shortest sampling interval of one second~\cite{wattsup_api}). However, the cumulative energy measurement (cumulative watt-hours) is performed  by the device continuously, by sampling the wattage measure 1000 times per second and then integrating the data to obtain cumulative energy usage, leading to higher accuracy. The device was connected to the computer running the workload via USB port and was controlled by automation scripts described in Section~\ref{sec:auto}. Cumulative energy measurement readings were taken before and after execution of a given workload. 

For measuring the system statistics (such as CPU and I/O load) we used Sysstat package \cite{sysstat}. The system measurements were taken asynchronously at one second interval during execution of the workload. Once a given workload execution was complete, the results were aggregated using summary statistics.

\subsubsection{Baseline Measurement}\label{sec:baseline}

As mentioned in Section~\ref{sec:testbed}, no other workloads were executed concurrently to ours. However, even though we tried to minimize the number of daemons ran by the OS, some of them have to be functional to ensure robustness of the OS. In addition, idle database engine and Sysstat package gathering system statistics may consume various amounts of energy.   

To quantify the impact of such  processes on energy consumption of the system, we measure energy consumption of the computer over 24 hour interval at 10 minutes increments (144 observations in total). The measurements were repeated for three different setups: 
\begin{enumerate}
    \item OS alone,
    \item OS with MySQL v.5.6.20 installed and running (no connections to the database were made), 
    \item OS with MySQL v.5.6.20 and Sysstat collecting system attributes at 1 second interval.
\end{enumerate}
The energy values were then converted to (average) power values to ``standardize'' the data. The conversion was done by dividing the amount of energy consumed by the time spent. 

Results of the measurements are given in Table~\ref{tab:baseline}. To assess variability of baseline energy and power consumption, we compute coefficient of variation as:
\begin{equation}\label{eq:coef-of-var}
    \sigma_v = \sigma / \mu, 
\end{equation}
where  $\mu$ and $\sigma$ are population mean and population standard deviation of the 144 observations for a given setup. The closer $\sigma_v$ to  zero -- the lower the variation. As we can see, OS alone consumes almost constant amount of energy through the day: $\sigma_v = 0.7\%$. MySQL consumes small amount of energy in idle state: extra 0.1W on top of the power consumed by OS alone. Systat consumes extra 0.2 W (or $0.3\% \gets 0.2/64.2$) of power. However, the consumption throughout the day remains almost constant: $\sigma_v = 0.5\%$. This lack of variability is very desirable in our case, as the baseline will not affect our analysis (as discussed in Section~\ref{sec:analysis}).

\begin{table}[t]
    \centering
    \begin{tabular}{@{}lrr@{}}
    \toprule
     \multicolumn{1}{c}{Baseline setup} & \multicolumn{1}{c}{Mean Energy Consumption (Wh)} & \multicolumn{1}{c}{Mean Power Consumption (W)}\\
        &  \multicolumn{1}{c}{ $\pm$ Coefficient of Variation} & \multicolumn{1}{c}{ $\pm$ Coefficient of Variation}\\
    \midrule
        OS alone                    &  $10.7 \pm 0.7\%$ &  $63.9 \pm 0.7\%$ \\
        OS with MySQL v.5.6.20      &  $10.7 \pm 0.7\%$ &  $64.0 \pm 0.7\%$ \\
        OS with MySQL v.5.6.20      &  $10.7 \pm 0.5\%$ &  $64.2 \pm 0.5\%$ \\
         \qquad and Sysstat         &                   & \\
    \bottomrule  
    \end{tabular}
    \caption{Baseline energy and power measurements.}
    \label{tab:baseline}
\end{table}

\subsection{ Software Metrics Extraction }\label{sec:softm}
We gathered the source code for each of the MySQL versions under study by downloading them from the original MySQL website \cite{mysql}. Then we created scripts to calculate the software code metrics for each MySQL version. In particular, we used CLOC tool \cite{cloc} to extract size metric ``total number of physical lines of code (without comments) in a given release'', denoted (LOC), and churn metric ``total number of lines of code changed in the the current release in comparison with the previous release'', denoted (LOCC). We also computed two complexity metrics---``traditional cyclomatic complexity''  (TCC) and ``modified cyclomatic complexity''  (MCC)---using the PMCCABE tool \cite{pmccabe}.  

Before computing the source code metrics, we eliminated a number of source code files. First, we eliminated source code files of test cases, since the code from them is not included in the production binaries of the database engine. Second, we removed source code not written in C and C++, as they are excluded from the production binaries.

\subsection{ Analysis of the data }\label{sec:analysis}
We analyzed the results (for a given product configuration) for the aggregate of all 22 SQL statements (queries)  as one unit of work, simulating a single analytic workload, as discussed in Sections~\ref{sec:workload} and \ref{sec:auto}. Energy, time, system's statistics, and software metrics data were stored in SQLite database~\cite{SQLite}.  R \cite{r, r_rsqlite} scripts (which obtained data from the SQLite database) were used to perform the analysis and produce tables and figures (discussed in Section~\ref{sec:results}).

We calculated the Pearson correlation coefficient\footnote{The Pearson correlation coefficient is a method to measure the linear relation (dependence) between any two variables $X$ and $Y$. We chose linear regression to analyze our data; therefore, we chose the Pearson correlation over the Spearman correlation, because Pearson is designed to match the sign and magnitude of a linear regression slope \cite{christine2004statistics}.} between all the variables used in our experiments (such as energy consumed and time spent, energy consumed and lines of code changed, and time spent and modified code complexity). 

We used the correlation data to answer the research questions, as explained in subsequent sections. We adopt the mapping between a value of correlation coefficient and the strength of correlation as per~\cite{christine2004statistics}. The mapping between the strength of correlation and a value of correlation coefficient is as follows:
\begin{enumerate*}[label={\alph*)}]
\item \textit{Perfect}: 1,
\item \textit{Strong}: 0.7-0.9,
\item \textit{Moderate}: 0.4-0.6,
\item \textit{Weak}: 0.1-0.3, and
\item \textit{Zero}: 0.
\end{enumerate*}

Note that Pearson correlation coefficient, denoted by $r$, is invariant to linear transformations; i.e., given two vectors $x$ and $y$, and a constant $c$, $r(x + c, y) = r(x,y)$. Based on the analysis in Section~\ref{sec:baseline}, our baseline energy consumption is almost constant. Therefore, our correlation analysis --- involving energy consumption as one of the variables --- will yield almost identical values of $r$ with baseline energy included or excluded from the energy consumption variable. In our case, we kept baseline energy included, as it represent the total amount of energy spent to complete an experiment.


\section{Results of Experiments}\label{sec:results}
This section is structured as follows.  The effect of system state on energy consumption is given in Section~\ref{sec:energy_vs_system}.  The relation between energy consumed and time spent is described in Section~\ref{sec:time_vs_energy}. System statistics analysis of the results of experiments, needed to answer research questions RQ1 and RQ2, are provided in Sections~\ref{sec:rq1_exp} and \ref{sec:rq2_exp}, respectively. Finally, the answers to RQ1 and RQ2 are provided and discussed in   Section~\ref{sec:discussion}.

\subsection{Energy consumption vs. System Statistics}\label{sec:energy_vs_system}
In this subsection we will analyze the relation between system statistics and energy consumption for all the experiments, independent of the release number. This will help us understand general constraints of the system under study. The discussion on per-release evolution is given in Section~\ref{sec:rq1_exp}-onward.

\begin{figure}
  
\end{figure}

\begin{figure}[!ht]
\centering
  \begin{subfigure}[b]{0.65\textwidth}
    \includegraphics[width=\textwidth]{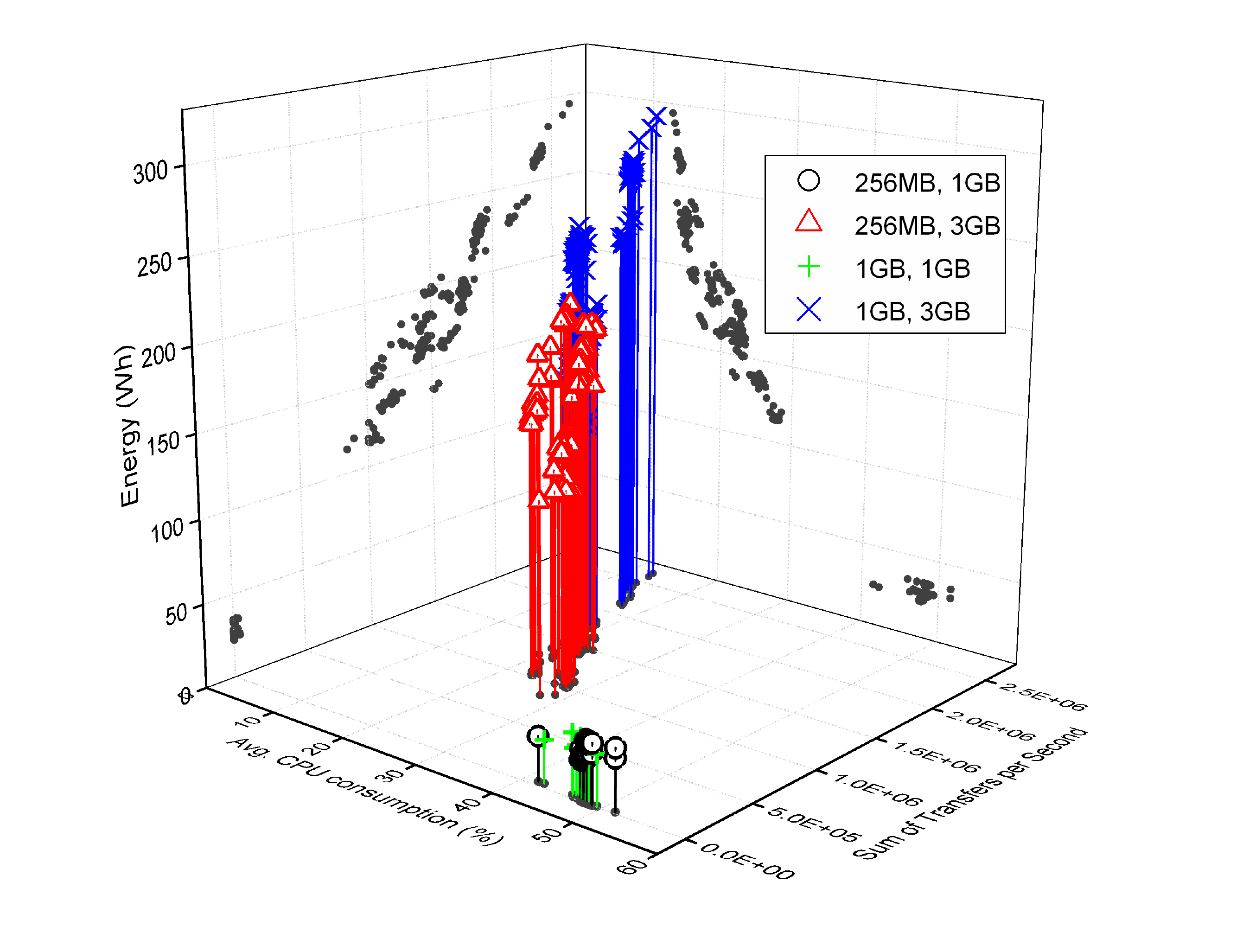}
    \caption{MySQL MyISAM}
    \label{fig:energy_vs_cpu_and_tps_myisam}
  \end{subfigure}
  \begin{subfigure}[b]{0.65\textwidth}
    \includegraphics[width=\textwidth]{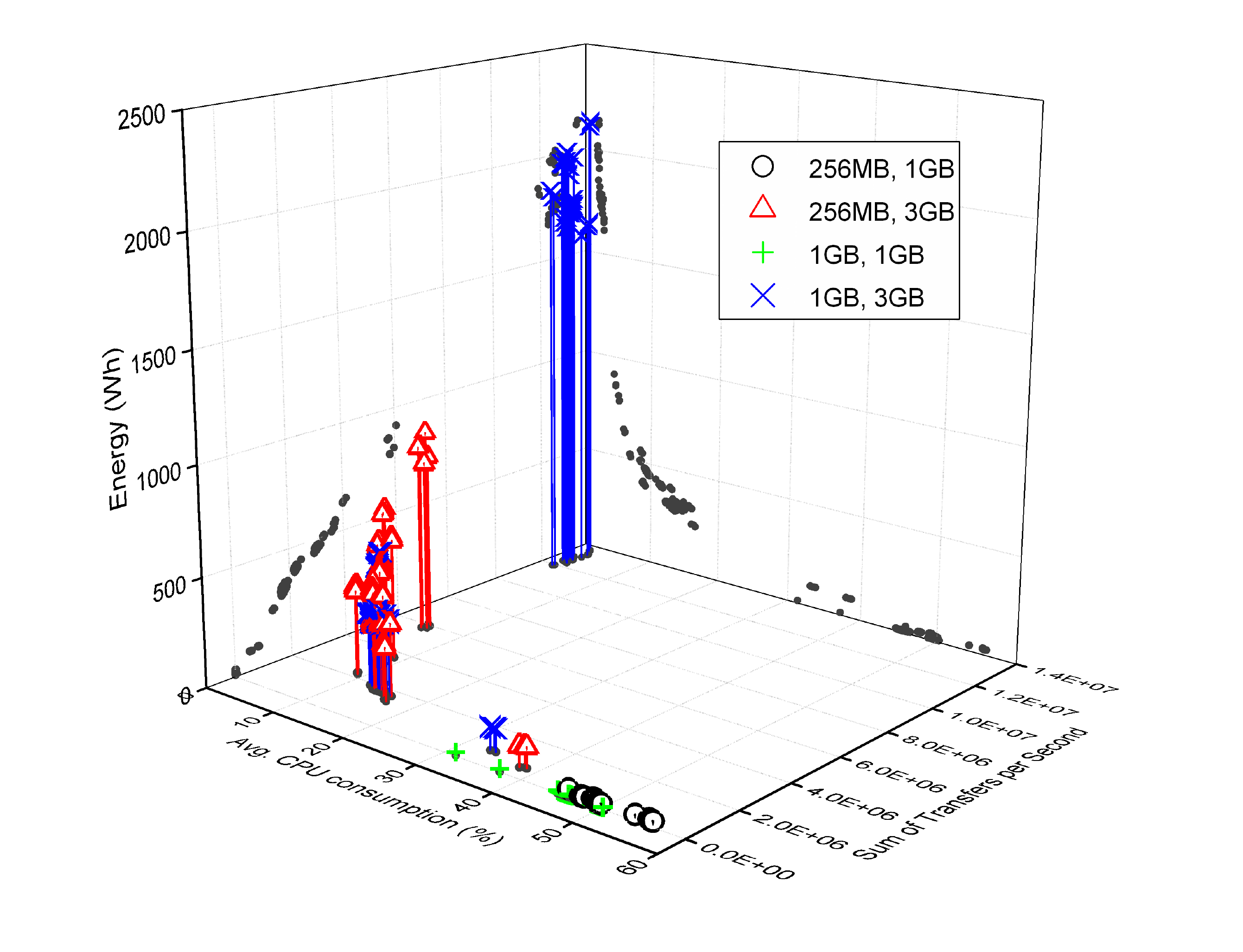}
    \caption{MySQL InnoDB}
    \label{fig:energy_vs_cpu_and_tps_innodb}
  \end{subfigure}
  
    \caption{Relation between CPU utilization, hard drive utilization, and energy consumed. Each data point represents a single workload execution. Different point types denote different setups (in term of memory buffer size and the amount of raw data) shown on the legend and summarized in Table~\ref{tab:experiments}. Vertical lines from the points are projections from data points to horizontal plane. Two-dimensional projections of CPU utilization vs. energy consumed and CPU utilization vs. hard drive utilization are given in Figure~\ref{fig:energy_vs_cpu_and_tps_2d}. }
    \label{fig:energy_vs_cpu_and_tps}
\end{figure}

\begin{figure}[!ht]
  \begin{subfigure}[b]{0.5\textwidth}
    \includegraphics[width=\textwidth]{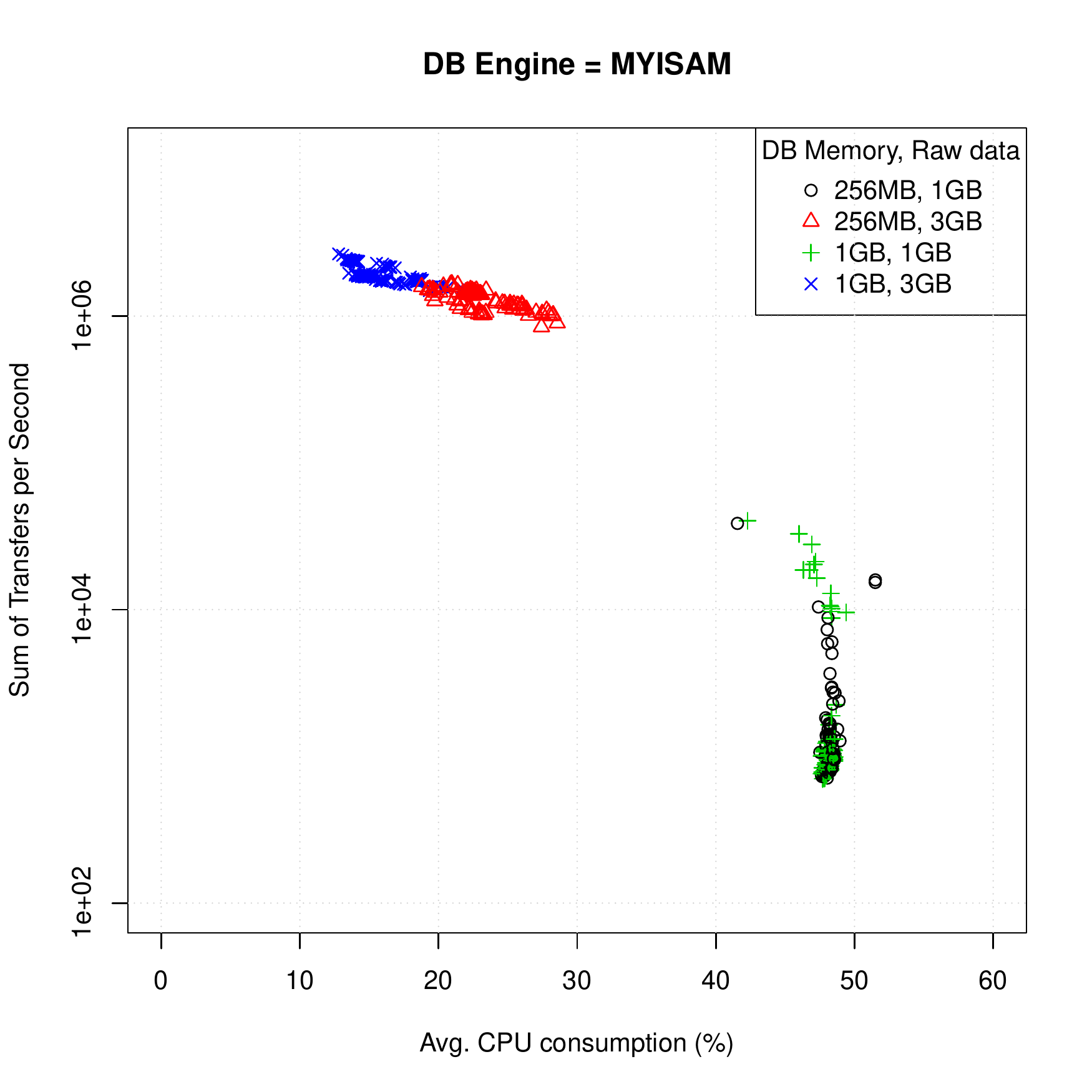}
    \caption{MySQL MyISAM: CPU vs. hard drive \\ utilization}
    \label{fig:cpu_and_tps_myisam}
  \end{subfigure}
  \begin{subfigure}[b]{0.5\textwidth}
    \includegraphics[width=\textwidth]{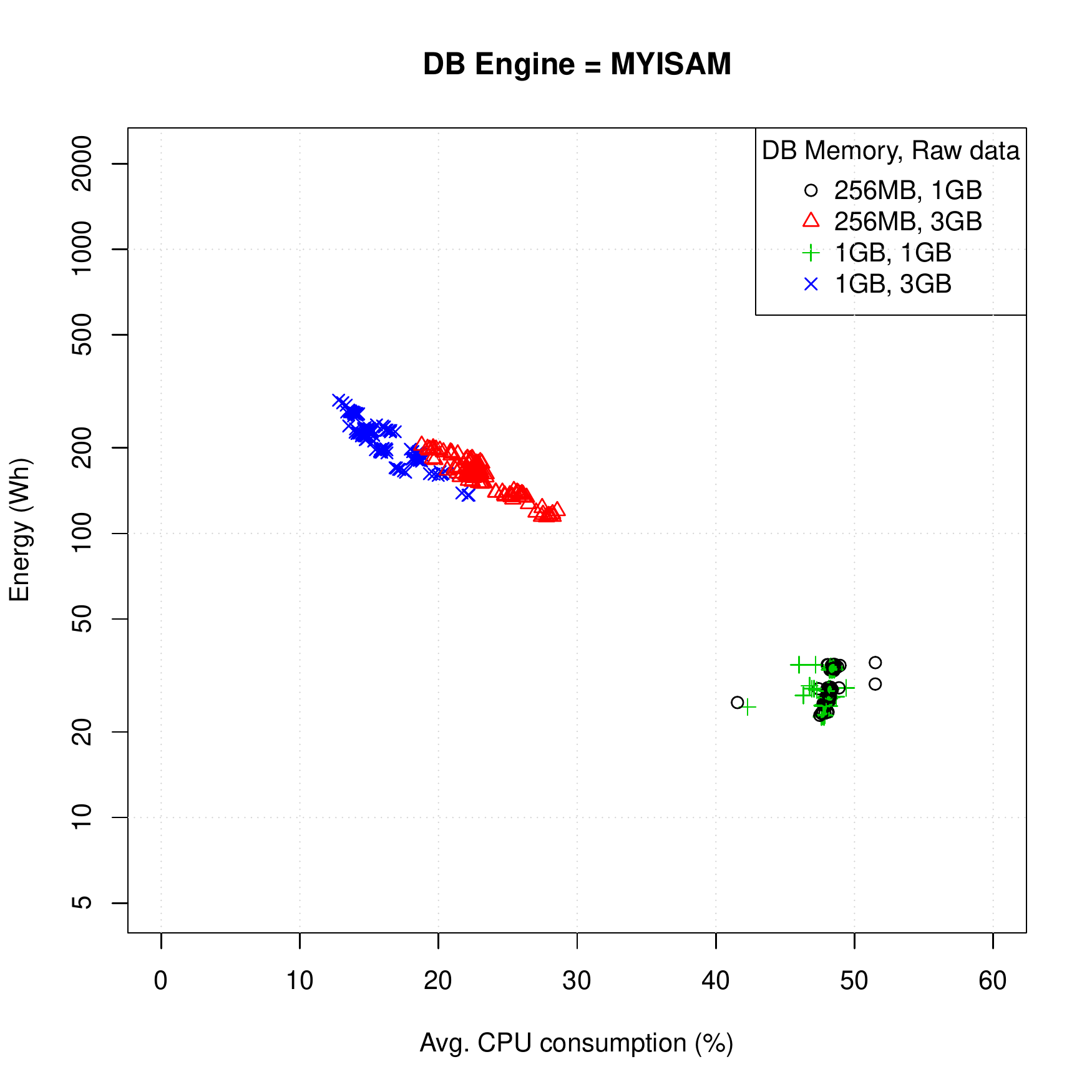}
    \caption{MySQL MyISAM: CPU utilization vs. energy consumed}
    \label{fig:cpu_and_energy_myisam}
  \end{subfigure}

  \begin{subfigure}[b]{0.5\textwidth}
    \includegraphics[width=\textwidth]{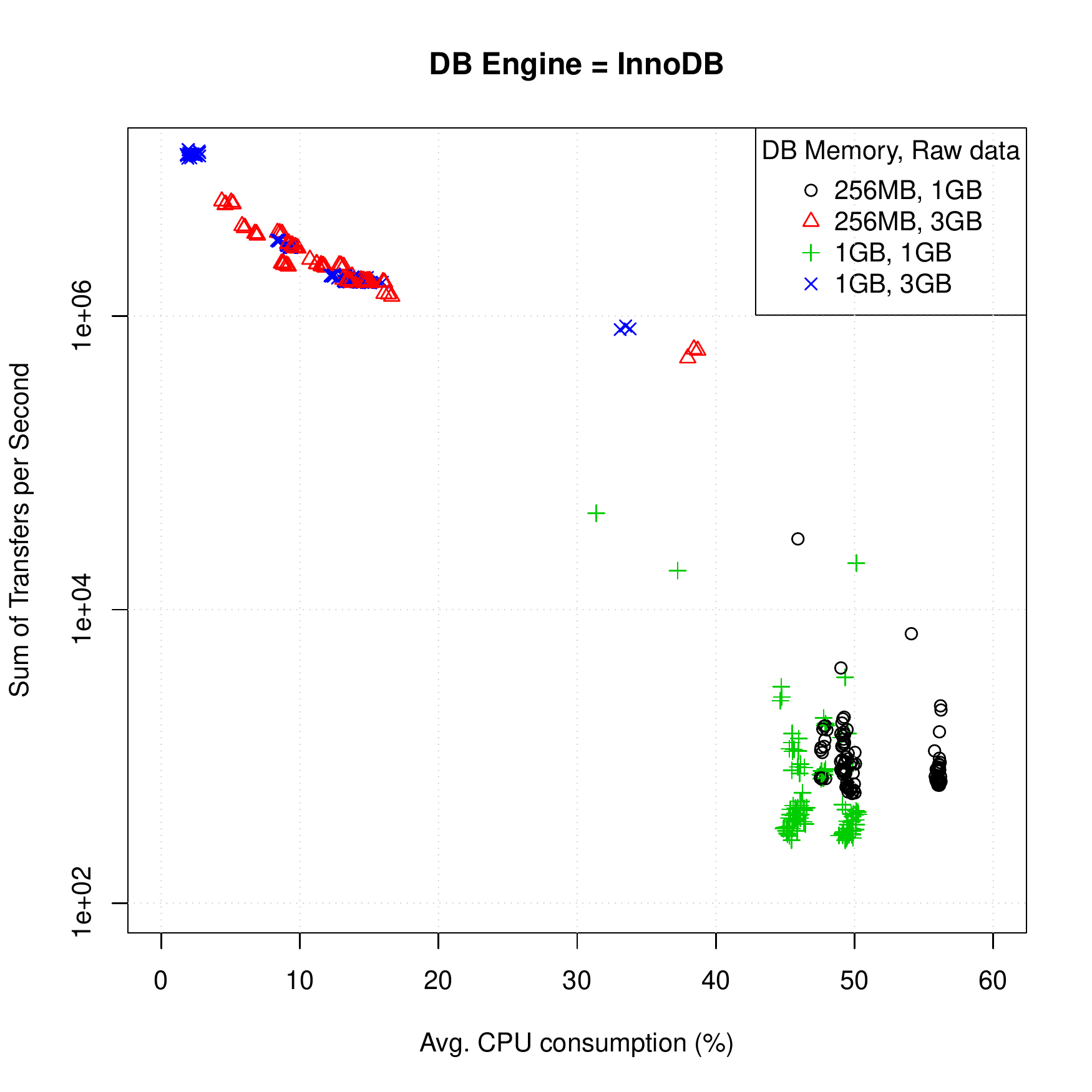}
    \caption{MySQL InnoDB: CPU vs. hard drive  utilization}
    \label{fig:cpu_and_tps_innodb}
  \end{subfigure}
  \begin{subfigure}[b]{0.5\textwidth}
    \includegraphics[width=\textwidth]{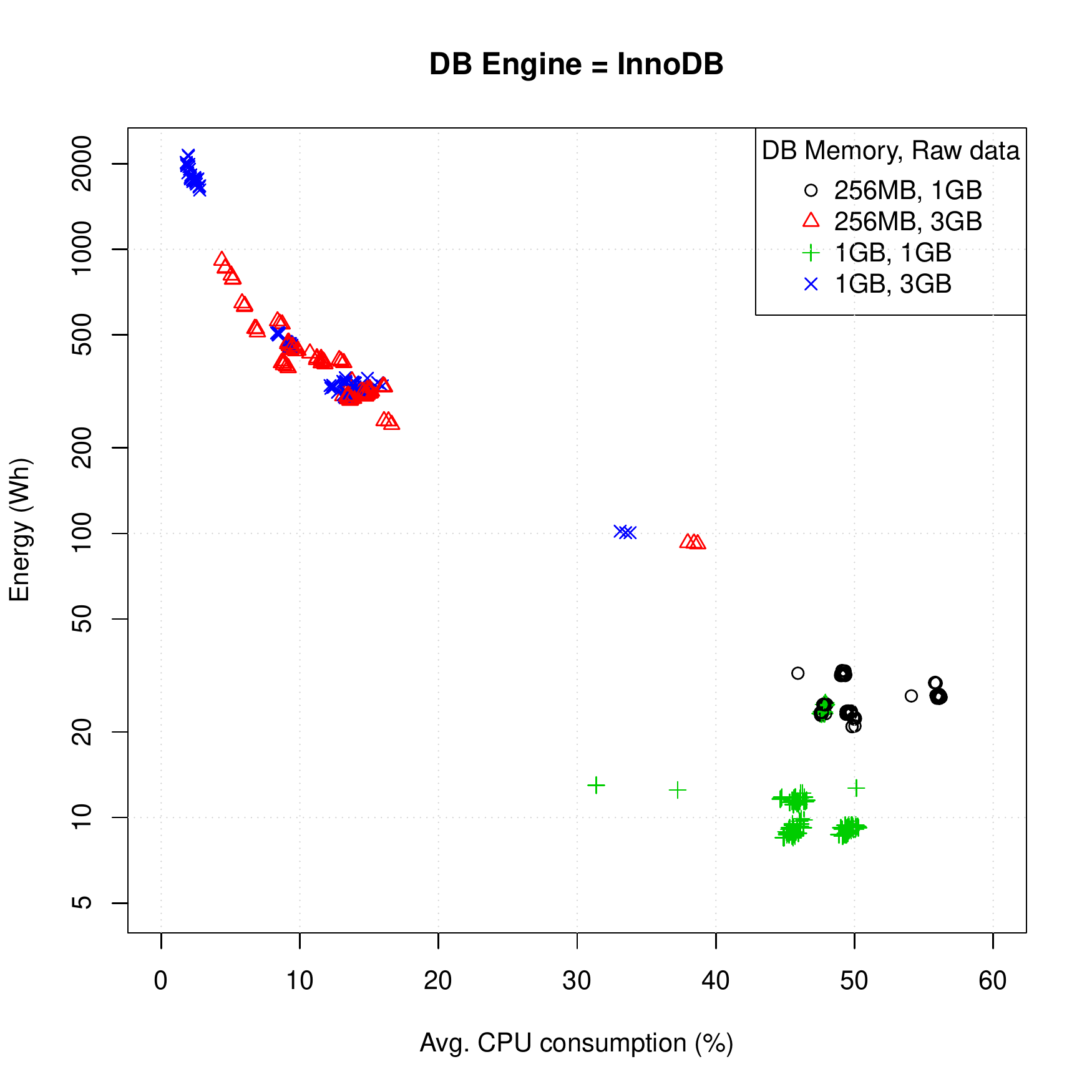}
    \caption{MySQL InnoDB: CPU utilization vs. energy consumed}
    \label{fig:cpu_and_energy_innodb}
  \end{subfigure}

    \caption{Relation between CPU utilization and hard drive utilization, and CPU utilization and energy consumed. The plots represent 2-dimensional projections of Figure~\ref{fig:energy_vs_cpu_and_tps}. Each data point represents a single workload execution. Different point types denote different setups (in term of memory buffer size and the amount of raw data) shown on the legend and summarized in Table~\ref{tab:experiments}.}
    \label{fig:energy_vs_cpu_and_tps_2d}
\end{figure}

Figure~\ref{fig:energy_vs_cpu_and_tps} and \ref{fig:energy_vs_cpu_and_tps_2d} show\footnote{A discriminating reader may notice counterintuitive results: providing more memory to a database engine does not yield expected performance improvement. We discuss this phenomenon in Section~\ref{sec:discussion:rq1}.} the relation between energy consumption, CPU utilization, and I/O load. We compute CPU utilization and I/O load as follows.

CPU utilization is computed as a sum of CPU utilization that occurred while executing at user level, user level with nice priority, and at the system level \cite{sysstat}. The CPU utilization data in this figure is computed by averaging out per-second data gathered by Sysstat. The higher the number -- the more utilized the CPU is. Note that our computer has two CPUs. Therefore, 50\% utilization means full load of one CPU. 

Transfers per second shows the number of I/O requests (of indeterminate size) to the hard drive. The higher the number the more data active our reads and writes to/from the hard drive. The transfers per second data in this figure is computed by summing up per second data collected by Sysstat.

As shown in Figures~\ref{fig:energy_vs_cpu_and_tps}, \ref{fig:cpu_and_tps_myisam}, and \ref{fig:cpu_and_tps_innodb}, the more I/O operations one has to do, the more CPU has to idle waiting for the I/O operations to complete. This leads to increased time spent and energy consumption, as the CPU idles, waiting for the data to be read from (and written to) the hard drive. 

As seen from Figure~\ref{fig:energy_vs_cpu_and_tps}, \ref{fig:cpu_and_tps_myisam}, and \ref{fig:cpu_and_tps_innodb}, the setup with 1GB of raw data being loaded into the database requires minimal amount of I/O operations, because the whole database is cached into memory. Once the data are cached -- all operations happen in memory, no access to the hard drive is required. In this case the workload often becomes CPU-bound, as the architecture of both engines (for the releases under study) cannot effectively parallelise processing of a single query. The setup with 3GB of raw data being loaded into the database is I/O-bound: the database engine cannot load all the data into memory and has to wait for I/O operations to complete.

InnoDB engine has a number of outliers with very high energy consumption ($> 500$Wh) for Experiments 6 and 8 (where 3GB of raw data are loaded into the database), as seen in Figures~\ref{fig:energy_vs_cpu_and_tps_innodb}, \ref{fig:cpu_and_tps_innodb}, \ref{fig:cpu_and_energy_innodb} and Figures~\ref{fig:exp7-bp}, \ref{fig:exp8-bp}. These outliers have a very high number of I/O operations associated with them (sum of transfers per second increases from $\approx 2.0 \times 10^6$ to $\approx 1.2 \times 10^7$ ). This large number of I/O operations can be explained by suboptimal decisions made by query optimizer. For example, it can decide to do a full table scan (i.e., read the content of every row in a table) rather than relying on indexes of a given table. Typically, this is considered a performance defect. Such defects are quickly noticed and reported by users (``My query took 10 minutes on a previous release; and now, after engine upgrade, it takes 10 hours!'') and are, usually, fixed quickly by the database engine developers. 

\subsection{ Execution Time vs. Energy Consumption }\label{sec:time_vs_energy}
\begin{figure}[t]
  \begin{subfigure}[b]{0.5\textwidth}
    \includegraphics[width=\textwidth]{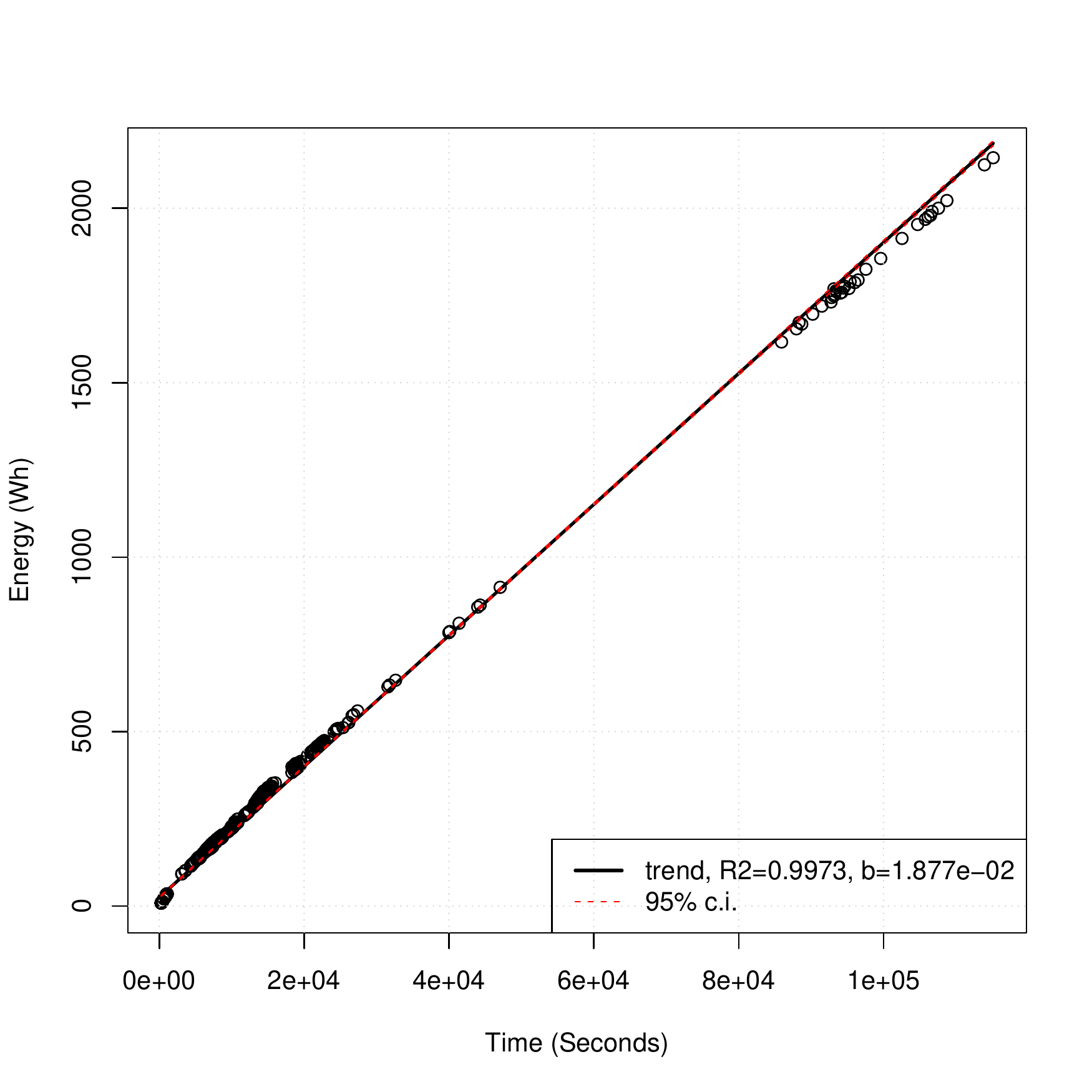}
    \caption{Time vs. Energy}
    \label{fig:time_vs_energy}
  \end{subfigure}
  \begin{subfigure}[b]{0.5\textwidth}
    \includegraphics[width=\textwidth]{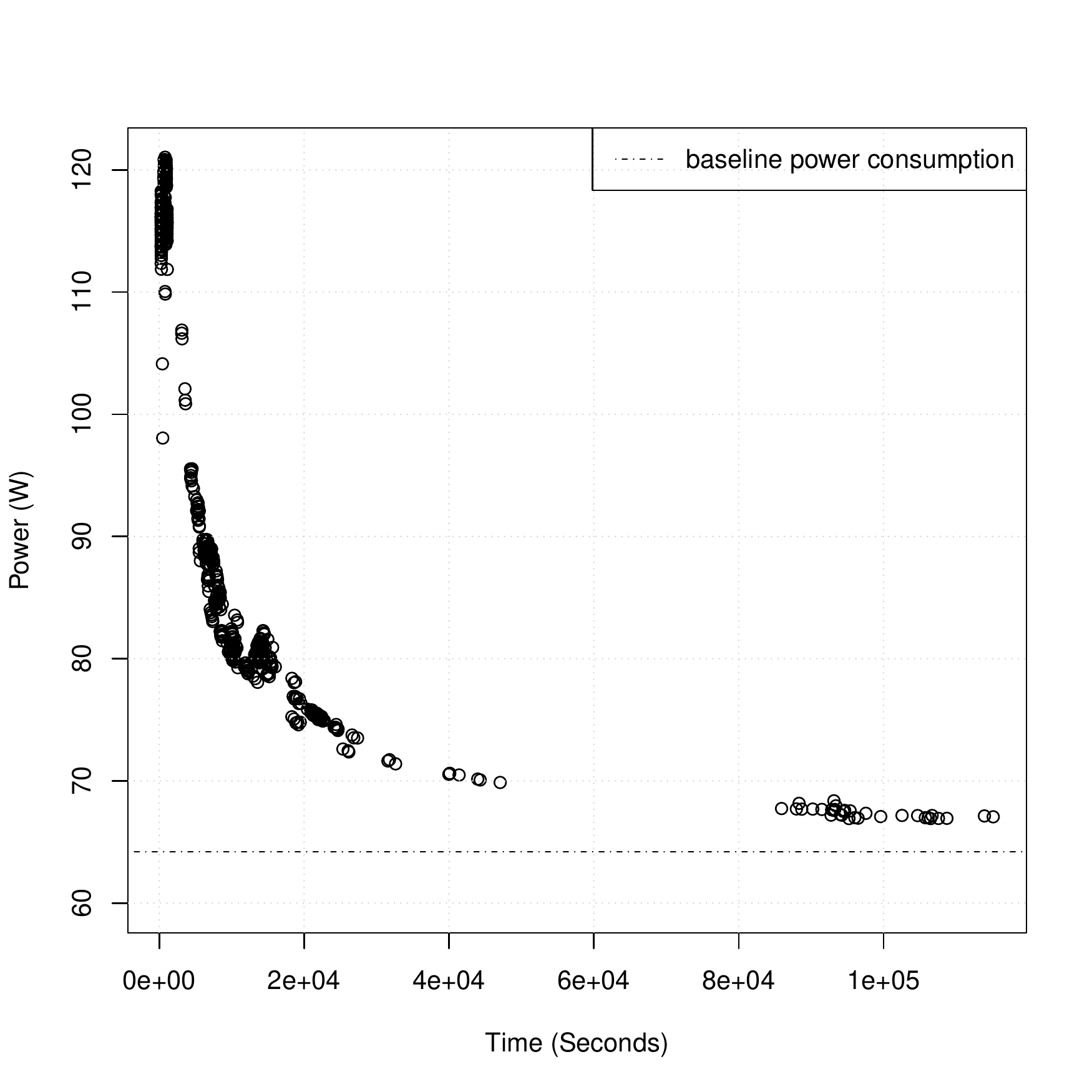}
    \caption{Time vs. Average Power}
    \label{fig:power_vs_energy}
  \end{subfigure}
  
    \caption{Relation between time spent and energy (average power) consumed for all experiments. A data point represents the time and energy (average power) data gathered for a given run of an experiment. The solid line depicts the trend line obtained using linear regression, and the dotted lines (that are very close to the solid line) show the 95\% confidence interval of the trend line. Dash-dotted horizontal line on Figure~\ref{fig:power_vs_energy} represents baseline average power consumption. }
    \label{fig:time_vs_energy_and_power}
\end{figure}

Figure~\ref{fig:time_vs_energy} shows the relation between energy consumption and execution time for all eight experiments. We can see that the strong positive linear relation between these two variables: coefficient of determination, $R^2 = 0.9975$ and the slope value, represented by $b$ in Figure~\ref{fig:time_vs_energy}, is positive. The Pearson correlation coefficient between these two variables is equal to $0.9986$, signifying an almost perfect correlation (as defined in Section~\ref{sec:softm}). 

This strong correlation comes from a trivial fact: the more time computer spends running the workload, the more energy it consumes. But what is the root cause of this observation? To get a better understanding of the situation, we plot time spent vs. average power consumed in Figure~\ref{fig:power_vs_energy}. The average power consumed is computed by dividing the amount of energy consumed by the time spent. From hereon, for the sake of brevity, we will use the term `power' instead of  `average power'. As the figure shows, the more time we spend -- the less power we consume. This can be explained by differences in the amount of power consumed by CPU and HDD. 

Both CPU and HDD require different amount of power while idling or under load. However, the amount of power (in absolute numbers) is vastly different: our HDD consume 5.6W while idling and 6.0W under load~\cite{hdd_spec}; each of our two CPUs, on the other hand, has thermal design power\footnote{Thermal design power is not equivalent to power consumed by the CPU. It gives the amount of power that CPU dissipates while being fully active. However, this value give us an understanding of the magnitude of power consumption.} of 84W~\cite{cpu_spec}. 

The higher the CPU utilization is -- the higher its power consumption. From Figure~\ref{fig:energy_vs_cpu_and_tps} we know that experiments which consumed the least amount of energy (and thus finished in the shortest period of time) had high CPU utilization. The figure also shows that  these workloads had almost no I/O activity, since all the data was loaded into memory. CPU did not have to idle waiting for I/O operations to complete and was able to maintain high utilization rate. These experiments correspond to data points in the bottom-right corner of Figure~\ref{fig:power_vs_energy}, where power consumption was in the range of 110-121W. 

As mentioned above, HDD consumes significantly smaller amount of energy than CPU (compare 6W with 84W). This becomes important for experiments causing high I/O utilization. Hard drive power consumption will remain at $\approx$6W, while CPU utilization will drop significantly (as shown in Figure~\ref{fig:energy_vs_cpu_and_tps}), as CPU idles waiting for I/O operations to complete. The cases of experiments with idling CPU  are represented by data points in top-left corner of Figure~\ref{fig:power_vs_energy}, where power consumption was $\approx$67W, which is quite close to the baseline power consumption of 64.2W. Even though power consumption in these cases is low, the energy consumption is high, since we have to integrate low power consumption value over prolonged time interval (when the CPU waits for I/O operations to complete). Similar behaviour was observed in the past with IBM DB2 database management system~\cite{koccak2013impact, miranskyy2013save}.

Keep in mind that each experiment yields the same result, the only difference is in the amount of ``work'' needed to obtain this result. To summarize, the experiments that consumed the most amount of time and energy are the I/O-bound ones, requiring high number of I/O operations.

\subsection{ RQ1: Experiments }\label{sec:rq1_exp}
In this section, we will analyze the results of our experiments to answer \textbf{RQ1}: `How does the energy consumption and execution time of a database engine change as the product matures (from release to release)?'

\subsubsection{Description of Equations, Tables, and Figures} \label{sec:rq1_descr}
In this subsection we describe tables (along with formulas needed to populate them) and figures that are used in subsequent Sections~\ref{sec:rq1_myisam} and \ref{sec:rq1_innodb}.

Figures~\ref{fig:energy_vs_version_exp_1-4} and \ref{fig:energy_vs_version_exp_5-8} show box plots of the energy consumed in MySQL MyISAM and InnoDB versions, respectively. Each figure contains four subplots -- one per experiment (listed in Table~\ref{tab:experiments}) for a given engine.  Note that we represent the data using box plots, since each experiment was conducted three times to ensure reproducibility.

Similarly, Figures~\ref{fig:time_vs_version_exp_1-4} and \ref{fig:time_vs_version_exp_5-8} contain box plots of the energy consumed in MySQL MyISAM and InnoDB versions, respectively. Because energy consumption and execution time are correlated almost perfectly, as was shown in Section~\ref{sec:time_vs_energy}, energy- and time- related pairs of figures are visually similar (Figures~\ref{fig:energy_vs_version_exp_1-4} and \ref{fig:time_vs_version_exp_1-4} for MyISAM and Figures~\ref{fig:energy_vs_version_exp_5-8} and \ref{fig:time_vs_version_exp_5-8} for InnoDB).

By eyeballing the plots, we can  observe trends in the data. To formally confirm our observations, we plot linear trend lines (computed using linear regression). We map release names to natural numbers in order to compute linear regression. The first minor release v.5.0.15 is mapped to 1, the second minor release v.5.0.16 is mapped to 2, etc. The $p$-values of the trend lines  for all but one product configuration are less than $0.001$. The remaining configuration (Experiment 8: InnoDB engine with 1GB of memory and 3GB of raw data shown in Figures~\ref{fig:exp8-bp} and \ref{fig:exp8-et} ) has a $p$-value of $\approx 0.05$ .

In subplots, solid line depicts a trend line obtained using linear regression; the dashed lines show the 95\% confidence interval of the trend line. The dotted lines represent a trend line (obtained using linear regression) per major release, and the dot-dashed lines show the 95\% confidence interval of these trend lines per major release. Vertical long-dash lines show boundaries between major releases. 

Figures contain a lot of information; in order to summarize this information, we construct the following tables. Table~\ref{tab:energy} summarizes the energy consumption for the MyISAM and InnoDB engine: Experiments \#1-4 in the table were conducted on MyISAM, Experiments \#5-8 -- on InnoDB. In this table, we show the average energy consumption of a minor release (average of three runs) that consumes the least amount of energy within a set of minor releases belonging to a given major release. Formally, for a given major release, the energy reading displayed in the tables is computed as follows:
\begin{equation}\label{eq:energy}
\min \left ( \frac{r_{1,1}+r_{1,2}+r_{1,3}}{3}, \frac{r_{2,1}+r_{2,2}+r_{2,3}}{3}, \cdots, \frac{r_{10,1}+r_{10,2}+r_{10,3}}{3} \right ),
\end{equation}
where $r_{i,j}$  represents energy consumption for the $j$-th run of the $i$-th minor release belonging to the major release of interest. The average of three runs per minor release in Eq.~\ref{eq:energy} is taken to minimize the measurement error and to reduce the effect of outliers for a given minor release; then the minimum amount of energy from the ten average readings is chosen. Furthermore, the relative percentage of difference between each two adjacent releases is calculated using the following formula:

\begin{equation}\label{eq:rel_diff}
\frac{\operatorname{new\_release\_value} - \operatorname{old\_release\_value}}{  \operatorname{old\_release\_value} } \times 100.
\end{equation}

To assess variability of minor releases for a given major release, we use Equation~\ref{eq:coef-of-var}, 
where  $\mu$ and $\sigma$ are population mean and population standard deviation of ten $(r_{i,1}+r_{i,2}+r_{i,3})/3$ terms.  The $\sigma_v$ values, shown in Table~\ref{tab:coefficient_of_variation}, portray the amount of variability in relation to $\mu$. The closer the $\sigma_v$ to $0$ -- the lower the variation.

The timing results per major release given in Table~\ref{tab:time} are computed using a formula that is structurally similar to that of Eq.~\ref{eq:energy}, as shown in Eq.~\ref{eq:time}: 
\begin{equation}\label{eq:time}
\min \left ( \frac{t_{1,1}+t_{1,2}+t_{1,3}}{3}, \frac{t_{2,1}+t_{2,2}+t_{2,3}}{3}, \cdots, \frac{t_{10,1}+t_{10,2}+t_{10,3}}{3} \right ),
\end{equation}
where $t_{i,j}$ represents workload execution time for the $j$-th run of the $i$-th minor release belonging to the major release of interest. The relative percentage of difference between each two adjacent releases is computed using Eq.~\ref{eq:rel_diff}. As in the case of Table~\ref{tab:energy}, Experiments \#1-4 in the table were conducted on MyISAM, Experiments \#5-8 -- on InnoDB. Given that energy consumption and execution time are correlated almost perfectly (see Section~\ref{sec:time_vs_energy} for details), timing results are similar to energy ones.

\subsubsection{ MyISAM } \label{sec:rq1_myisam}

We performed four Experiments (\#1--4) on the MyISAM engine. The analysis of the per-run data, collected during these four experiments, suggests that both the energy consumption and execution time of the MyISAM engine increased as the engine matured. In essence, on average, the newer major releases are slower and less green than the older ones. 

Table~\ref{tab:energy} summarizes the energy consumption for the MyISAM engine. The oldest release, v.5.0, is the greenest (of all major releases), because it consumed the least amount of energy to execute the same reference workload. On the other hand, the newest release, 5.6, is the brownest (least green), because it consumed the largest amount of energy, as shown in Table~\ref{tab:energy}. 

All four experiments showed a consistent increase in energy consumption between any two major consecutive releases (all the difference values in Tables~\ref{tab:energy} are positive, representing this increase). For example, Experiment 1 shows that release v.5.1 is 14.12\% less efficient than v.5.0, release v.5.5 is 5.30\% less efficient than release v.5.1, and release v.5.6 is 18.71\% less efficient than release v.5.5 -- see Table~\ref{tab:energy} for details. 

Figure~\ref{fig:energy_vs_version_exp_1-4} shows that, overall, the energy consumption increased as MySQL matured. By eyeballing the plots, we can see that this trend is clearly pronounced as we move from one previous major release to the next one. This is also confirmed by linear models: they explains most of the data variability (based on $R^2$ values): in the Experiments \#1-3 from 83\% to 87\% of variability is explained (see Figures \ref{fig:exp1-bp}, \ref{fig:exp2-bp}, and \ref{fig:exp3-bp}),  while in the case of Experiment \#4 only 61\% of variability is explained (see Figures~\ref{fig:exp4-bp}). 

The increase in variability for Experiments \#4 and, to a degree, for Experiment \#2 (see Figures~\ref{fig:exp4-bp} and \ref{fig:exp2-bp}, respectively) can be explained by the fact that Experiments \#1 and 3 process 1GB of raw data, all of which can be fitted in memory. However, in Experiments \#2 and 4, we cannot fit all 3GB of raw data into memory. Database memory manager and OS file caching mechanism need to derive sophisticated probabilistic strategies to efficiently load the data to memory from the hard drive (e.g., by predicting which data pages will soon be needed and proactively loading them into memory  \cite{mysql_prefetching}). We will discuss this observation for both database engines in Section~\ref{sec:discussion}.

As we can see from Table~\ref{tab:coefficient_of_variation}, the variation of energy consumption within three major releases v.5.1, v.5.5, and v.5.6 remains low ($ \le 0.02$). The results are further confirmed by eyeballing Figure~\ref{fig:energy_vs_version_exp_1-4} and examining relatively flat trend lines for these releases. However, the variation for v.5.0 is higher (between 0.04 and 0.08). Moreover, the trend line in Figure~\ref{fig:exp4-bp} suggests that energy consumption increases from the earliest minor release v.5.0.15 to the latest minor release v.5.0.96 under study. We conjecture that this anomalous behaviour may be related to the fact that v.5.0 is the first major release in a series of v.5.x releases; it could have had ``infantile diseases'' that were outgrown in subsequent releases. 

The timing results per major release are given in Table~\ref{tab:time} and Figure~\ref{fig:time_vs_version_exp_1-4}. Since energy consumption and execution time are correlated almost perfectly, the timing results are very close to the energy consumption results. In other words, execution time increases as the engine matures, with the v.5.0 release being the fastest and v.5.6 being the slowest. We conjecture that this behaviour of the MyISAM engine can be explained by the fact that the inclusion of additional functionality to the MyISAM engine in each subsequent release requires additional computational resources, leading to an increase of time and energy consumption. 

\begin{table}[t]
    \centering \small
    \ra{1.3}
    \begin{tabular}{@{}lrrrrrrrr@{}}
    \toprule
     & \multicolumn{4}{c}{Minimum energy (Eq. \ref{eq:energy}) consumed   } & & \multicolumn{3}{c}{ Energy consumption: relative   } \\
     & \multicolumn{4}{c}{by a minor release  } & & \multicolumn{3}{c}{  difference (Eq. \ref{eq:rel_diff}) between  } \\
     & \multicolumn{4}{c}{for a given major release (Wh)  } & & \multicolumn{3}{c}{  previous and current  release   } \\
    \cmidrule{2-5} \cmidrule{7-9}
    Major release & v.5.0  & v.5.1  & v.5.5  & v.5.6  & & v.5.1  & v.5.5  & v.5.6 \\
    \midrule
    1: MyISAM, 256, 1 & \best{23.13}  & 26.40  & 27.80  & \worst{33.00}  & &  14\% & 5\% & 19\% \\
    2: MyISAM, 256, 3  & \best{114.87} & 155.53 & 164.10 & \worst{182.77} & & 35\% & 6\% & 11\% \\
    3: MyISAM, 1024, 1  & \best{22.83}  & 26.40  & 27.97  & \worst{33.17}  & &  16\% & 6\% & 19\% \\
    4: MyISAM, 1024, 3  & \best{156.17} & 208.40 & 212.10 & \worst{252.03} & & 33\% & 2\% & 19\% \\
    \addlinespace
    5: InnoDB, 256, 1 & 23.00  & 26.30  & \best{20.93}  & \worst{31.67}  & & 14\%  & -20\% & 51\% \\
    6: InnoDB, 256, 3  & \best{92.30}  & 310.43 & 246.40 & \worst{448.07} & & 236\% & -21\% & 82\% \\
    7: InnoDB, 1024, 1  & 8.63   & 9.03   & \best{8.57}   & \worst{11.30}  & & 5\%   & -5\%  & 32\% \\
    8: InnoDB, 1024, 3  & \best{101.00} & 318.07 & 307.10 & \worst{473.77} & & 215\% & -3\%  & 54\% \\
    \bottomrule
    \end{tabular}
    \caption{Minimum energy consumption data per major release with relative percentages of difference between each pair of adjacent releases. \textbf{Bold} text highlights the greenest major release for a given experiment; \textit{italic} text -- the brownest (least green) one. The content of the first column is constructed based on the following template: `Experiment \#: DB Engine Name, Memory Buffer Size (MB), Raw Data Size~(GB)'.}
    \label{tab:energy}
\end{table}

\begin{table}[t]
    \centering \small
    \ra{1.3}
    \begin{tabular}{@{}lrrrrrrrr@{}}
    \toprule
    & \multicolumn{4}{c}{Minimum execution time (Eq. \ref{eq:time}) taken   } &  & \multicolumn{3}{c}{ Execution time: relative   } \\
    & \multicolumn{4}{c}{by a minor release  } & &  \multicolumn{3}{c}{  difference (Eq. \ref{eq:rel_diff}) between  } \\
    & \multicolumn{4}{c}{for a given major release (Hr)  } & &  \multicolumn{3}{c}{  previous and current  release   } \\
    \cmidrule{2-5} \cmidrule{7-9}
    Major release & v.5.0  & v.5.1  & v.5.5  & v.5.6  & &  v.5.1  & v.5.5  & v.5.6 \\
    \midrule
    1: MyISAM, 256, 1   & \best{0.20}  & 0.23  & 0.24  & \worst{0.29}  & &  15\% & 6\% & 20\% \\
    2: MyISAM, 256, 3   & \best{1.21}  & 1.75  & 1.83  & \worst{2.09}  & &  45\% & 5\% & 14\% \\
    3: MyISAM, 1024, 1  & \best{0.20}  & 0.23  & 0.24  & \worst{0.29}  & &  16\% & 6\% & 19\% \\
    4: MyISAM, 1024, 3  & \best{1.83}  & 2.55  & 2.58  & \worst{3.14}  & &  39\% & 1\% & 22\% \\
    \addlinespace
    5: InnoDB, 256, 1 & 0.20 & 0.22 & \best{0.18} & \worst{0.27} & &  11\%  & -20\% & 52\%  \\
    6: InnoDB, 256, 3  & \best{0.87} & 3.83 & 2.96 & \worst{5.92} & &  342\% & -23\% & 100\% \\
    7: InnoDB, 1024, 1  & 0.07480 & 0.08 & \best{0.07479} & \worst{0.10} & &  5\%   & -5\%  & 29\%  \\
    8: InnoDB, 1024, 3  & \best{1.00} & 3.96 & 3.86 & \worst{6.30} & &  297\% & -2\%  & 63\%  \\
    \bottomrule
    \end{tabular}
    \caption{Minimum execution time data per major release with the relative percentage of difference between each pair of adjacent releases. \textbf{Bold} text highlights the fastest major release for a given experiment; \textit{italic} text -- the slowest one. The content of the first column is constructed based on the following template: `Experiment \#: DB Engine Name, Memory Buffer Size (MB), Raw Data Size~(GB)'.}
    
    \label{tab:time}
\end{table}

\begin{table}[t]
    \centering
    \begin{tabular}{@{}lrrrrrrrrr@{}}
    \toprule
            & \multicolumn{4}{c}{MyISAM   } & & \multicolumn{4}{c}{ InnoDB   } \\
    \cmidrule{2-5} \cmidrule{7-10}
    Experiment \#  & 1 & 2 & 3 & 4 && 5 & 6 & 7 & 8 \\
    \midrule
    v.5.0                       & 0.04         & 0.07         & 0.04         & 0.08         && 0.08         & 0.27         & 0.49         & 1.03         \\
    v.5.1                       & 0.01         & 0.02         & 0.003         & 0.01        && 0.01         & 0.22         & 0.04         & 0.91         \\
    v.5.5                       & 0.01         & 0.01         & 0.01         & 0.01         && 0.04         & 0.24         & 0.04         & 0.97         \\
    v.5.6                       & 0.02         & 0.02         & 0.01         & 0.01         && 0.01         & 0.30         & 0.02         & 0.77         \\
    \bottomrule
    \end{tabular}
    \caption{Coefficient of variation in energy consumption (Eq. \ref{eq:coef-of-var}) of minor releases for a given major release. }
    \label{tab:coefficient_of_variation}
\end{table}

\subsubsection{InnoDB}\label{sec:rq1_innodb}

We conducted four Experiments (\#5--8) on the InnoDB engine. The summary of the energy consumption (computed using Eq.~\ref{eq:energy}) for these experiments is given in Table~\ref{tab:energy} along with the relative percentage of difference between each pair of adjacent releases, calculated using Eq.~\ref{eq:rel_diff}. We can see that the results for the InnoDB engine are less monotone (in comparison with the MyISAM engine). Examination of Table~\ref{tab:energy} reveals that energy consumption increased from v.5.0 to v.5.1, decreased from v.5.1 to v.5.5, and increased again from v.5.5 to v.5.6 for all four experiments. For example, for Experiment 5 energy consumption increased by 14.53\% from v.5.0 to v.5.1, decreased by 5.51\% from v.5.1 to v.5.5, and, finally, increased by 19.73\% from v.5.5 to v.5.6. This non-linear dynamics is confirmed by low levels of variability explained by linear trend lines in Figure~\ref{fig:energy_vs_version_exp_5-8}: $R^2$ for the lines ranges between 0.03 and 0.31.

As in the case of MyISAM, the least green major release is the most recent one (namely, v.5.6). However, we obtained different results for the greenest engine: in the case of Experiments \#6 and 8, the greenest version is the oldest one (v.5.0), while, in the case of the remaining Experiments, \#5 and 7, the greenest engine is the intermediate v.5.5. Experiments \#5 and 7 cache all the data in the memory (using 1GB raw data size); Experiments \#6 and 8 have to read the data from the hard drive (using 3GB raw data size). This implies that the functionality of v.5.5 may be better suited for handling an I/O-intensive workload. 

Variability between minor releases of InnoDB is also significantly higher than in the case of MyISAM, as shown in Table~\ref{tab:coefficient_of_variation}: e.g., compare $\sigma_v=0.08$ for v.5.0 in Experiment 4 with  $\sigma_v=1.03$ for v.5.0 in Experiment 8. The variability for Experiments \#5 and 7 is $<0.08$ (with the exception of v.5.0 with $\sigma_v=0.49$); the variability for Experiments \#6 and 8 is significantly higher with $\sigma_v$ ranging between $0.22$ and $1.03$. 

Moreover, examination of Figure~\ref{fig:energy_vs_version_exp_5-8} reveals that variation between minor releases can be significant (as we already discussed in Section~\ref{sec:energy_vs_system}): e.g., in Experiment \#8, based on Figure \ref{fig:exp8-bp}, v.5.6.15 consumed $\approx 260\%$ more energy than its predecessor v.5.6.14 or its successor v.5.6.16 (average energy readings are 1748.93, 479.56, and 476.56, respectively). IT personnel maintaining a database running on InnoDB engine should be mindful of this fact: an untested upgrade to a new release may lead to significant increase in energy consumption (and performance degradation, since these two variables are strongly correlated).

Table~\ref{tab:time} and Figure~\ref{fig:time_vs_version_exp_5-8} summarize the execution time for Experiments \#5--8.  Due to correlation of time and energy consumption the timing data mirrors the energy data: v.5.6 is the slowest in all four experiments. The fastest release in Experiments \#6 and 8 is v.5.0, while v.5.5 is the fastest for Experiments \#5 and 7.

We will discuss the difference in behaviour between the MyISAM and InnoDB engines in Section~\ref{sec:discussion}.

\subsection{ RQ2: Experiments }\label{sec:rq2_exp}

In this section, we will show the analysis of the data focusing on \textbf{RQ2}: `Which software metrics reflect energy consumption and execution time'? In order to answer RQ2, we analyzed the relations between energy consumption and software metrics (namely, LOC, LOCC, TCC and MCC) in all eight experiments. In Section~\ref{sec:rq2_descr} we describe equations, figures, and tables that are used in analyzing the relations in Section~\ref{sec:rq2_rel}. 

\subsubsection{Description of Equations, Tables, and Figures} \label{sec:rq2_descr}
We provide exploratory analysis of the data, namely, the distributions of the software metrics and correlations among the software metrics in Figures~\ref{fig:stats_code_metrics} and \ref{fig:cor_code_metrics}, respectively. We list Pearson correlation coefficient values between energy consumption (or time spent) and various software metrics in each experiment in Tables~\ref{tab:cor_energy_vs_metrics} and \ref{tab:cor_time_vs_metrics}, respectively.

We introduce two additional response variables: change in energy consumption and change in time spent. They are defined as: 
\begin{equation} \label{eq:e_delta}
\operatorname{avg\_energy\_consumed\_by\_minor\_release\_N} - \operatorname{avg\_energy\_consumed\_by\_minor\_release\_N-1 }
\end{equation}
and
\begin{equation} \label{eq:t_delta}
\operatorname{avg\_time\_spent\_by\_minor\_release\_N} - \operatorname{avg\_time\_spent \_by\_minor\_release\_N-1}.
\end{equation}
The variables represent the difference between energy consumed (or time spent) by a given release and previous release. We will use these variables to  identify relations between energy consumed (or time spent) and software metrics.

Table~\ref{tab:cor_energy_dt_vs_metrics} shows the Pearson correlation values between changes in energy consumed (Eq.~\ref{eq:e_delta}) and the LOCC metric. 
Table~\ref{tab:cor_time_dt_vs_metrics} gives the Pearson correlation values between changes in time spent (Eq.~\ref{eq:t_delta}) and the LOCC metric. Given the perfect correlation between energy consumed and time spent, the resulting numbers in these two tables are very close: the difference between the correlation values is less than 0.03.

\subsubsection{The Relations} \label{sec:rq2_rel} 
\begin{figure}[t]
  \begin{subfigure}[b]{0.5\textwidth}
    \includegraphics[width=\textwidth]{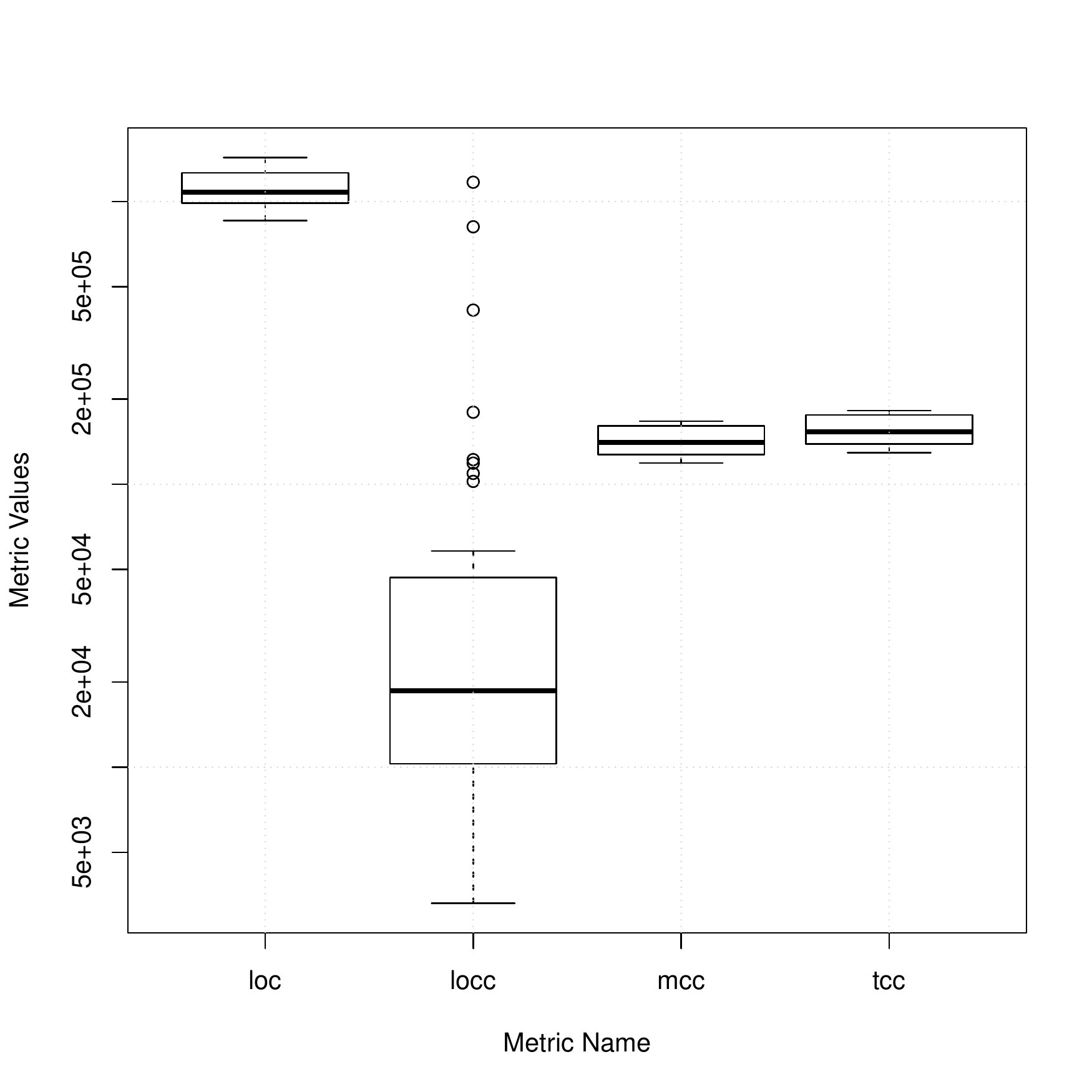}
    \caption{Distributions of the software metrics.}
    \label{fig:stats_code_metrics}
  \end{subfigure}
  \begin{subfigure}[b]{0.5\textwidth}
    \includegraphics[width=\textwidth]{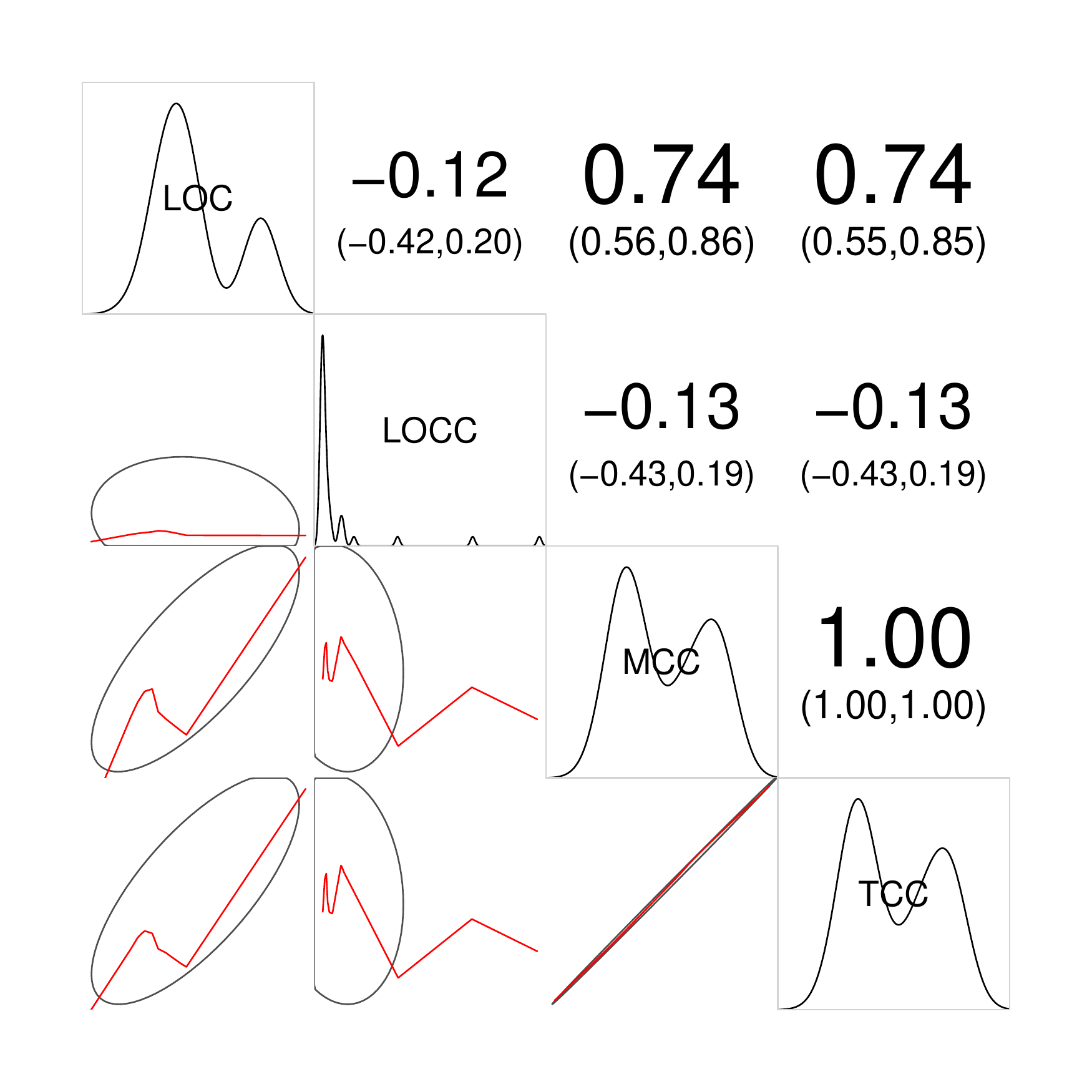}
    \caption{Correlogram of the software metrics.}
    \label{fig:cor_code_metrics}
  \end{subfigure}
  
    \caption{Left panel contains a box plot showing distributions of the software metrics. Right panel presents correlogram visualizing correlation matrix for  the software metrics. Diagonal shows metric name and distribution of the metric. Lower triangle region shows confidence ellipse and smoothed line. Upper triangle region shows Pearson’s correlation coefficient $r$ and, in brackets, confidence interval for $r$.}
\end{figure}

The distributions of the metrics in Figure~\ref{fig:stats_code_metrics} show that LOCC has higher variability in comparison with the other three metrics. By examining correlation among the software metrics (see Figure~\ref{fig:cor_code_metrics}), we found that LOC is strongly correlated with MCC and TCC. This behaviour has been observed in other software products in the past \cite{hindle2012green, Hindle2015, lind1989experimental}. MCC and TCC are perfectly correlated, by construction, since their formulas are similar. LOCC is weakly correlated with the other variables.

We investigated the relation between energy consumption (time spent) and each software metric graphically, using corresponding scatter plots, and numerically, by analyzing the trend lines obtained using linear regression and correlation between variables. 

By examining correlations between energy consumption (or time spent) and various software metrics in each experiment (given in Tables~\ref{tab:cor_energy_vs_metrics} and \ref{tab:cor_time_vs_metrics}), we see that the results are identical due to an almost perfect correlation between energy and time. 

These tables show that there exists a moderate to strong positive correlation between consumed energy (or time spent) and codebase size, measured in LOC, for Experiments \#1-6 for both database engines: MyISAM and InnoDB. However, for Experiments \#7 and 8 on InnoDB engine, the correlation between energy consumed (or time spent) ranges between none to weak. This may be explained by the high variability of the results in these two Experiments (as per Table~\ref{tab:coefficient_of_variation} and Figure~\ref{fig:energy_vs_version_exp_5-8}), ``masking'' the correlation. 

In order to test this hypothesis, we removed the outliers (i.e., those data points whose energy consumption and performance were anomalously high, probably due to performance defects, as discussed in Section~\ref{sec:energy_vs_system}) and recalculated the correlation for Experiments \#5--8 (see values in brackets in Tables~\ref{tab:cor_energy_vs_metrics} and \ref{tab:cor_time_vs_metrics}). As we can see, removal of the outliers does yield strong correlation between consumed energy (or time spent) and LOC for Experiment \#7. However, for Experiment \#8 the correlation remains weak.

In the case of MyISAM, there is no correlation between LOCC and consumed energy (or time spent); while in case of InnoDB there is none to weak correlation (with the exception of Experiment 8, where correlation is moderate). The exception can be explained by high variability in the data and is, probably, a statistical artifact. The removal of the outliers does not change the picture significantly: the correlation for Experiments \#5-7 remains weak to none, and for Experiment \#8 the correlation drops from moderate to weak.

The correlation between energy consumed (or time spent) and code complexity metrics (MCC and TCC) ranges from none to strong, suggesting that it cannot be used as a consistent predictor of consumed energy. Removal of the outliers, increases correlation strength to the `weak to strong' range, but does not affect the conclusion. Peculiarly, an examination of Figure~\ref{fig:cor_code_metrics} suggests that the relation between LOC and complexity metrics (MCC and TCC) is strong. Nevertheless, MCC and TCC are less reliable predictors than LOC. 

We also computed the correlation  between changes in energy consumed (Eq.~\ref{eq:e_delta}) --- or time spent (Eq.~\ref{eq:t_delta}) --- and the software metrics LOC, MCC, and TCC. We do not show Pearson correlation coefficient values for the sake of brevity, but the correlation for all experiments was none to weak. 

The correlation data in Tables~\ref{tab:cor_energy_dt_vs_metrics} and \ref{tab:cor_time_dt_vs_metrics} show that the change in energy consumption (or change in time spent) between releases is correlated with the code churn, measured by LOCC; the correlation for the MyISAM engine ranges between moderate and strong, while the correlation for InnoDB engine ranges between none and weak. Removal of the outliers increase InnoDB correlation range from weak to moderate. This suggests that changes in energy consumption or time has some correlation with the amount of changes to the code; however, the size of code attribute has stronger relation.

\begin{table}[t]
    \centering
    \begin{tabular}{@{}lrrrrrrrrr@{}}
    \toprule
            & \multicolumn{4}{c}{MyISAM   } && \multicolumn{4}{c}{ InnoDB   } \\
    \cmidrule{2-5} \cmidrule{7-10}
    Experiment \#  & 1 & 2 & 3 & 4 && 5 & 6 & 7 & 8 \\
    \midrule
    LOC     & 0.93         & 0.73         & 0.92         & 0.69         && 0.74         & 0.62         & -0.06        & 0.12 \\
            &              &              &              &              && (0.76)         & (0.60)         & (0.89)         & (0.15) \\
    \addlinespace
    LOCC    & 0.00         & 0.02         & 0.01         & 0.02         && 0.03         & 0.15         & -0.01        & 0.43 \\
            &              &              &              &              && (0.03)         & (-0.22)        & (-0.03)        & (0.30) \\
    \addlinespace
    MCC     & 0.53         & 0.21         & 0.49         & 0.31         && 0.82         & 0.49         & 0.26         & 0.03 \\
            &              &              &              &              && (0.80)         & (0.52)         & (0.76)         & (0.35) \\
    \addlinespace
    TCC     & 0.51         & 0.19         & 0.47         & 0.29         && 0.80         & 0.48         & 0.28         & 0.03 \\
            &              &              &              &              && (0.78)         & (0.50)         & (0.75)         & (0.35) \\
    \bottomrule
    \end{tabular}
    \caption{Pearson correlation coefficient values between energy consumption and the software metrics. The values in brackets for Experiments \#5--8 show correlation coefficient values for dataset without outliers.}
    \label{tab:cor_energy_vs_metrics}
\end{table}

\begin{table}[t]
    \centering
    \begin{tabular}{@{}lrrrrrrrrr@{}}
    \toprule
            & \multicolumn{4}{c}{MyISAM   } && \multicolumn{4}{c}{ InnoDB   } \\
    \cmidrule{2-5} \cmidrule{7-10}
    Experiment \#  & 1 & 2 & 3 & 4 && 5 & 6 & 7 & 8 \\
    \midrule
    LOCC & 0.55 & 0.75 & 0.74 & 0.66 && 0.27 & 0.09 & 0.06 & -0.01 \\
    & & & &  &&(0.29)    &	(-0.27) &	(0.44)  &	(0.31) \\
    \bottomrule
    \end{tabular}
    \caption{Pearson correlation coefficient values between changes in energy consumption and LOCC. The values in brackets for Experiments \#5--8 show correlation coefficient values for dataset without outliers. }
    \label{tab:cor_energy_dt_vs_metrics}
\end{table}

\begin{table}[t]
    \centering
    \begin{tabular}{lrrrrrrrrr}
    \toprule
            & \multicolumn{4}{c}{MyISAM   } && \multicolumn{4}{c}{ InnoDB   } \\
    \cmidrule{2-5} \cmidrule{7-10}
    Experiment \#  & 1 & 2 & 3 & 4 && 5 & 6 & 7 & 8 \\
    \midrule
    LOC & 0.93  & 0.71 & 0.92 & 0.67 && 0.75 & 0.63 & -0.07 & 0.12 \\
    &              &              &              &              && (0.77)       & (0.62)       & (0.86)       & (0.16) \\
    \addlinespace
    LOCC & -0.01 & 0.03 & 0.01 & 0.02 && 0.03 & 0.15 & -0.02 & 0.43 \\
    &              &              &              &              && (0.03)       & (-0.23)      & (-0.03)      & (0.30) \\
    \addlinespace
    MCC  & 0.53  & 0.20 & 0.49 & 0.30 && 0.84 & 0.48 & 0.26  & 0.03 \\
    &              &              &              &              && (0.82)       & (0.52)       & (0.75)       & (0.34) \\
    \addlinespace
    TCC  & 0.51  & 0.18 & 0.47 & 0.29 && 0.82 & 0.47 & 0.27  & 0.02 \\
    &              &              &              &              && (0.81)       & (0.51)       & (0.74)       & (0.34) \\
    \bottomrule

    \end{tabular}
    \caption{Pearson correlation coefficient values between time spent and the software metrics. The values in brackets for Experiments \#5--8 show correlation coefficient values for dataset without outliers.}
    \label{tab:cor_time_vs_metrics}
\end{table}

\begin{table}[ht]
    \centering
    \begin{tabular}{lrrrrrrrrr}
    \toprule
            & \multicolumn{4}{c}{MyISAM   } && \multicolumn{4}{c}{ InnoDB   } \\
    \cmidrule{2-5} \cmidrule{7-10}
    Experiment \#  & 1 & 2 & 3 & 4 && 5 & 6 & 7 & 8 \\
    \midrule
    LOCC & 0.53 & 0.76 & 0.73 & 0.66 && 0.30 & 0.10 & 0.07 & -0.01 \\
    & & & & && (0.31) & (-0.28) &	(0.46) &	(0.31) \\
    \bottomrule
    \end{tabular}
    \caption{Pearson correlation coefficient values between changes in time spent and LOCC. The values in brackets for Experiments \#5--8 show correlation coefficient values for dataset without outliers. }
    \label{tab:cor_time_dt_vs_metrics}
\end{table}

\subsection{Discussion}\label{sec:discussion}

\subsubsection{ RQ1	}\label{sec:discussion:rq1} The answer to RQ1: `How does the energy consumption and execution time of a database engine change as the product matures (from release to release)?' is as follows. 

As shown in Tables~\ref{tab:energy} and \ref{tab:time} and Figure~\ref{fig:energy_vs_version_exp_1-4}, the overall energy consumption for MyISAM engine increases as the product matures, suggesting that the additional functionality added with every new release consumes additional resources. In the case of InnoDB (based on data in Tables~\ref{tab:energy} and \ref{tab:time} and Figure~\ref{fig:energy_vs_version_exp_5-8}), the earliest major release v.5.1 is the greenest in 50\% (2 out of 4) of the experiments, while the latest major release v.5.6 is the least green in 100\% (4 out of 4) experiments. In the case of the two remaining experiments, the greenest title is claimed by an intermediate release v.5.5. 

The results of execution time (performance) are similar to energy consumption results (due to the almost perfect correlation between energy consumed and time spent, see Section~\ref{sec:time_vs_energy}): in MyISAM case, the overall execution time increases as the product matures, i.e., newer releases are slower than the older ones. InnoDB execution time findings are identical to energy consumptions findings as well. 

This is different from the results of the experiments on the Firefox web-browser \cite{hindle2012green, Hindle2015}, where energy consumption decreased as the product matured. This suggests that, depending on the product and its domain, the results may vary. This difference can potentially be explained by the fact that the two products' (Firefox and MySQL) application domain, construction methods, and coding styles are different.


Let us discuss additional difference in behaviour between the MyISAM and InnoDB engines, shown in Sections~\ref{sec:rq1_myisam} and \ref{sec:rq1_innodb}.

\textbf{Greenness and performance}: based on Tables~\ref{tab:energy} and \ref{tab:time} data,  the results for greenest and fastest results of InnoDB are better than those of MyISAM. However, this rule is not universal. For example, a database user (whose data cannot fit into memory) may require some features present only in the latest version v.5.6 of MySQL. In this case, the user may have to select MyISAM engine instead of InnoDB, since it will be 61\% greener (182.77Wh vs. 473.77Wh) and 67\% faster (2.09Hr vs. 6.30Hr), see results of Experiments 2 and 8 in Tables~\ref{tab:energy} and \ref{tab:time} for details.

\textbf{Variation within major release}: based on the data from Table~\ref{tab:coefficient_of_variation} and Figures~\ref{fig:energy_vs_version_exp_1-4} and \ref{fig:energy_vs_version_exp_5-8}, MyISAM's results are less volatile than InnoDB ones. We conjecture that this can be explained by the fact that  MyISAM was designed to handle analytic workloads, while InnoDB was originally designed for transactional ones \cite{mysql_engines}. Since the developers have not focused on satisfying requirements of analytic workloads, the performance and energy consumption results for the current analytic workload are volatile. 

Peculiarly, before major release v.5.5, MYSAM was a default database engine for MySQL; starting from v.5.5, InnoDB became the default one \cite{mysql}. Database administrators interested in executing an analytic workload should take into account this difference while setting up their databases. This has to be taken into consideration during migration to a new version of the engine or an upgrade to a new release, since it may lead to a spike in energy consumption and performance degradation (as discussed in Section~\ref{sec:rq1_innodb}).

\textbf{Effect of Raw Data Size}: based on Tables~\ref{tab:energy} and \ref{tab:time}, we can see that in Experiments 1 vs. 5 and 3 vs. 7, dealing with 1GB of raw data that can fit into computer's memory, InnoDB engine is greener and faster than MyISAM engine for all releases. In the case of Experiments 2 vs. 6 and 4 vs. 8, dealing with 3GB of raw data that cannot fit into memory, MyISAM engine is greener and faster than InnoDB engine with one exception: InnoDB's early release v.5.0 (in Experiment 6) is greener and faster than MyISAM's one (in Experiment 2).

This suggests that InnoDB may be better suited for in-memory processing than MyISAM. However, for larger datasets that cannot fit into memory (which is more common for analytic workloads), MyISAM may be a better choice (as mentioned above, MyISAM was designed for analytic workloads).

\textbf{Effect of Memory Buffer}: Tables~\ref{tab:energy} and \ref{tab:time} show that in the case of in-memory processing (for Experiments \# 1 vs. 3 and 5 vs. 7, dealing with 1GB raw data) increase of the memory buffer from 256MB to 1024MB leads to marginal (less than 1\%) improvement in greenness and performance for MyISAM engine. In the case of InnoDB the improvement is significant (approximately 60\%). This can be explained by the fact that MyISAM has a rudimentary memory manager, relying mainly on OS mechanism for file caching, while InnoDB has a more sophisticated memory manager \cite{mysql_myisam, mysql_innodb}.

In the case of processing data that cannot fit into memory (experiments 2 vs. 4 and 6 vs. 8, dealing with 3GB of raw data) increase of the memory buffer leads to significant (29\% to 38\% for MyISAM and 2\% to 25\% for InnoDB) degradation of greenness and performance for both MyISAM and InnoDB engines. This can be explained by the fact that OS mechanism for file caching outperforms database engines memory manager for all major releases.

\subsubsection{ RQ2 }

The answer to RQ2: `Which software metrics reflect energy consumption and execution time?' is as follows. Consumed energy is governed mainly by the size of the code base. As shown in Section~\ref{sec:rq2_exp}, the code size LOC metric serves as a moderate to strong predictor of energy consumption for both database engines (with the exception of InnoDB's Experiment 8, where correlation is weak). The code churn (LOCC) and complexity (MCC and TCC) metrics results are weaker. This conclusion holds for predicting energy consumed as well as change in energy consumption. 

The results for execution time (performance) are similar (due to the almost perfect correlation between energy consumed and time spent, see Section~\ref{sec:time_vs_energy}): The time is governed mainly by LOC; LOCC, MCC and TCC have a lesser effect on performance. These results suggest that the amount of consumed energy and time spent are governed by the sheer volume of code to execute rather than the amount of changes introduced or code complexity. If we treat high energy consumption as a defect \cite{penzenstadler2014safety}, our results differ from the results seen for functional defects, where LOCC acts as a better predictor of defects than does LOC \cite{misirli2011different, miranskyy2014effect}. This result is also different from the findings of Hindle \cite{hindle2012green, Hindle2015}, who found that, in the case of the Firefox web browser, LOC is not correlated with power consumption.

Note that LOC cannot work as a predictor of energy consumption or time spent when a performance defect, causing major performance degradation, is present in the code base. However, such cases are rare (as we observed and discussed in Section~\ref{sec:energy_vs_system}).

\section{Threats to validity}\label{sec:threats}
Here we present threats to validity, classified as per~\cite{yin2013case, wohlin2012, Hindle2015}.

\textbf{Conclusion validity }: There are a number of threats related to reliability of measures. In order to avoid fluctuation of energy consumption due to \textit{fluctuation of ambient temperature} (higher temperature may lead to higher energy consumption by computer's hans), the computer was placed in thermostated environment of the data center. 

\textit{Baseline energy consumption may fluctuate over time.} In order to address this threat, we installed server version of the operating system with minimal amount of software preinstalled. We also performed extensive measurements of the baseline, making sure that it remains nearly constant over time.

\textbf{Construct validity}: To make \textit{comparison of different releases} representative, we chose a subset of ten minor releases from all the available major MySQL releases (v.5.0, v.5.1, v.5.5, and v.5.6), providing representative ``picture'' for each major release, helping to gain a broad and clear idea about MySQL's conduct.

\textit{The limitation in MySQL behaviour (response) if using a single configuration.} To mitigate this threat, we chose two different database storage engines types (MyISAM and InnoDB) in order to get a more general idea about the database engine's behaviour. Also we chose different key buffer sizes (256MB and 1024MB), which helped in examining the various situations for each database engine. Moreover we chose different raw data sizes (1GB and 3GB), which helped to gain a clear idea about the different responses of the database engine when the raw data was less than the available memory (all data can be cashed in the case of 1GB) or exceeded the available memory (in the case of 3GB). 

To ensure  \textit{consistency in the workload used}, we ran the same standard TPC-H workload to be sure that we had a consistent execution in all experiments. 

To maintain \textit{accuracy and precision of results} and reduce \textit{measurement error}, we ensured that all the required measurements (such as energy consumption and execution time) were calculated automatically. Moreover, each experiment was repeated three times, increasing accuracy and precision, and reducing measurement error.


\textbf{Internal validity}: In order to reduce the threat to validity related to \textit{human errors}, we automated the process of data gathering and analysis, reducing the risk of human error. We created Python \cite{python}  scripts for gathering code metrics from the source code of the database engine. We also created a Python script that profiled workload execution. The results of the experiments as well as the source code metrics were stored in a SQLite database. We then automatically generated the figures and correlation tables using R \cite{r, r_rsqlite} scripts (which accessed data from the SQLite \cite{SQLite} database). 

\textbf{External validity}: As described by Wieringa and Daneva~\cite{wieringa2015}, software engineering studies suffer from the variability of the real world, and the generalization problem cannot be solved completely. As they indicate, to build a theory we need to generalize to a theoretical population and have adequate knowledge of the architectural similarity relation that defines the theoretical population. We studied two database engines. The \textit{generalization to other database engines or other software products} is, obviously, not possible. However, the software under study represents a critical case \cite{yin2013case} of a relational database management system. In this study we do not aim build a theory, rather we would like to have a deeper understating of energy footprint in a case study. However, our experimental framework can be applied to other projects with well-designed and controlled experiments.

\section{Conclusions and Future Work}\label{sec:conclusions}

In this research our aim was to explore and have a deeper understanding of the impact of energy efficiency on database applications. We performed a case study, measuring the energy consumption and execution time of two MySQL database engines across 40 releases on a reference analytic workload TPC-H. To achieve this goal, we developed a framework to measure the energy consumption and execution time of a database workload by extracting software metrics from each release of the software product.

Answering RQ1---`How does the energy consumption and execution time of a database engine change as the product matures (from one release to another)?'---our study shows that the MySQL MyISAM engine becomes less green and less efficient as the product matures in all four experiments (leading to increased energy consumption and higher CO$_2$ emissions). In the case of MySQL InnoDB engine, the earliest major release is the greenest and fastest in 50\% (2 out of 4) of our experiments, while the latest major release is the least green and efficient in 100\% (4 out of 4) of our experiments. 

This is different from the results of the experiments on the Firefox web-browser \cite{hindle2012green, Hindle2015}, where energy consumption decreased as the product matured. This difference suggests that, depending on the product and its domain, energy consumption and execution time may ``evolve'' differently.

Answering RQ2---`Which software metrics reflect energy consumption and execution time?'---we show that consumed energy and performance are governed mainly by the size of the code base. The code size LOC metric serves as a moderate to strong predictor of energy consumption and performance for both database engines (with the exception of one InnoDB experiment). The smaller the code base, the greener and more efficient the database engine is. The code churn (LOCC) and complexity (MCC and TCC) metrics results have a lesser effect on energy consumption and performance. 

This implies that the amount of consumed energy and time spent are governed by the volume of the code to execute rather than the amount of changes introduced to the code base or the code complexity. If we treat high energy consumption as a defect (as per \cite{penzenstadler2014safety}), our results differ from the results seen for functional defects, where LOCC acts as a better predictor of defects than does LOC \cite{misirli2011different, miranskyy2014effect}. This result also differs from the finding of Hindle \cite{hindle2012green, Hindle2015}, who found that, in the case of the Firefox web browser, LOC is not correlated with power consumption.

LOC cannot work as a predictor of energy consumption or time spent when a performance defect, causing major performance degradation, is present in the code base. However, such cases are rare, as they are quickly exposed by the users (as observed and discussed in Section~\ref{sec:energy_vs_system}).

To summarize, our answers to the research questions suggest that, depending on the product and its domain, the results may vary. For example, the difference between Hindle's \cite{hindle2012green, Hindle2015} and our findings may be explained by the  differences in the application domain, construction methods, and coding styles of the products under study (Firefox and MySQL). 

Our findings may give insights to both practitioners and researchers. Database administrators may use our findings to select a green and fast release of the MySQL database engine. Developers of MySQL database engines may assess the greenness and performance of their product with the help of software metrics. The findings may also be of interest to researchers, as they lay a foundation for models predicting the greenness and performance of databases, which, in turn, would aid in developing green software. 

Going forward, we plan to expand our work by studying other relational, NoSQL, and NewSQL database engines as well as study other enterprise products, such as middleware servers. We also would like to collect requirements data to better understand the release content and its impact on energy consumption.

\appendix
\section{Test Automation: Algorithm} \label{sec:experiment_details}
In this appendix we provide pseudo code for automatic execution of test cases and gathering energy, time, and systems statistics for various releases and configurations of the database engine. The main procedure is given in Algorithm~\ref{alg:main}. The main procedure, in turn, calls procedure to setup test database (Algorithm~\ref{alg:setup}) and function to execute test case, also known as workload, (Algorithm~\ref{alg:wkld}). Note that we do not explicitly define procedures for installing, uninstalling, starting, stopping, and restarting a database engine, as these procedures are platform-specific.

Note that we recommend to rejuvenate~\cite{Cotroneo2014} energy meter on a regular basis via reboot of its software (see Alogrithm~\ref{alg:main}, line~\ref{lst:rejuv}), which can be done programmatically. We have noticed that after a few weeks of continuous log collections, the meter may start to behave erratically; rejuvenation fixes this issue.

\begin{algorithm}[ht]
\caption{Main test harness}\label{alg:main}
\begin{algorithmic}[1]
\Procedure{main\_harness}{}

    \State $db\_engines \gets $ [myisam, innodb] 
    \State $db\_engines\_versions \gets [ 5.0.15, 5.0.16, \ldots, 5.6.21 ]$ \Comment{ list of versions for each db engine }
    \State $db\_sizes \gets [1, 3]$ \Comment{size of raw data (in GB) to be loaded into test database}
    \State $mem\_cfgs \gets [256, 1024]$ \Comment{memory buffer size (in MB)}
    \State $test\_db \gets $ `your\_db\_name' \Comment{name of db under study}
    \State $workload  \gets $ a list of SQL statements \Comment{TPC-H workload}
    \State $runs \gets 3$ \Comment{run each workload three times}

   \ForEach {$db\_size \in db\_sizes$}
        \ForEach {$db\_engine \in db\_engines$}\        
            \State $upgrade\_type \gets$ `FreshInstall'
            \ForEach {$db\_version \in db\_engines\_versions$}
                \ForEach {$mem\_cfg \in mem\_cfgs$}
                    \State \Call{uninstall\_existing\_database\_engine}{} \textbf{if} exists
                    \State \Call{install\_database\_engine}{$db\_engine, db\_version$}
                    \State \Call{start\_database\_engine}{$db\_engine, db\_version, mem\_cfg$}
                    \State \Call{setup\_test\_db}{$test\_db, db\_size, upgrade\_type$}
                    \State $upgrade\_type \gets$ `Upgrade' 
                    \For{$run\_id \gets 1$ \textbf{to} $runs$}
    	          	    \State soft-reset energy meter \Comment{rejuvenate to increases robustness of the meter} \label{lst:rejuv}
    	          	    \State \Call{restart\_database\_engine}{$db\_engine, db\_version, mem\_cfg$}
    	                \State  {$[energy\_consumption, processing\_time, system\_stats] \gets$ \Call{execute\_workload}{$test\_db, workload$} }
    	                \State \Call{store}{ $db\_engine$, $db\_version$, $db\_size$, $mem\_cfg$, $run\_id$, $energy\_consumption$, $processing\_time$, $system\_stats$}\Comment{Save statistics} 
    	                \State \Call{stop\_database\_engine}{$db\_engine, db\_version$} \Comment{Cleanup}
    	                \State \Call{uninstall\_existing\_database\_engine}{} \Comment{Cleanup}
                    \EndFor
                \EndFor
            \EndFor
        \EndFor
   \EndFor
   \State do final cleanup: delete $test\_db$ and \Call{uninstall\_existing\_database\_engine}{}
\EndProcedure
\end{algorithmic}
\end{algorithm}

\begin{algorithm}[ht]
\caption{Setup test database}\label{alg:setup}
\begin{algorithmic}[1]

\Procedure{setup\_test\_db}{$db, db\_size, upgrade\_type$}
    \If {$upgrade\_type =$ 'FreshInstall'} \Comment{Mimics fresh install}
        \State delete $db$ if exists
        \State create new $db$
        \State create objects (tables, indices, etc.) in $db$
        \State populate objects with data based on the $db\_size$
    \ElsIf{upgrade\_type = 'Upgrade'} \Comment{Mimics upgrade from a previous release to a new release}
        \State run a script upgrading database  from a previous release \Comment{the script is specific to a given engine, e.g., \texttt{mysql\_upgrade}. Note that, typically, in the case of upgrade within major release (i.e. from one minor releases of a given major release to another one) no action is required.}
    \Else
        \State \Call{exit}{`Unknown upgrade type'}
    \EndIf
\EndProcedure    

\end{algorithmic}
\end{algorithm}

\begin{algorithm}[ht]
\caption{Execute workload}\label{alg:wkld}
\begin{algorithmic}[1]

\Function{execute\_workload}{$db, workload$}
    \State $start\_energy \gets$ current cumulative Wh value
    \State $start\_time \gets$  current time
    \State start asynchronous system stats gathering at 1 second interval
    \State connect to $db$
    \ForEach {$sql\_statement \in workload$}
	    \State  execute sql\_statement and get the recordset
	 \EndFor
    \State stop system stats gathering
    \State $consumed\_energy \gets$ current cumulative Wh value  $- start\_energy$
    \State $end\_time \gets$ current time  
    \State $consumed\_time \gets $ current time $ - start\_time $
    \State $system\_stats \gets$ get summary statistics on CPU and I/O utilization from Sysstat
    \State \textbf{return} $[consumed\_energy, consumed\_time, system\_stats]$

\EndFunction

\end{algorithmic}
\end{algorithm}
 
\section{Supplementary material for Results of Experiments}

\begin{figure}[ht]
  \begin{subfigure}[b]{0.5\textwidth}
    \includegraphics[width=\textwidth]{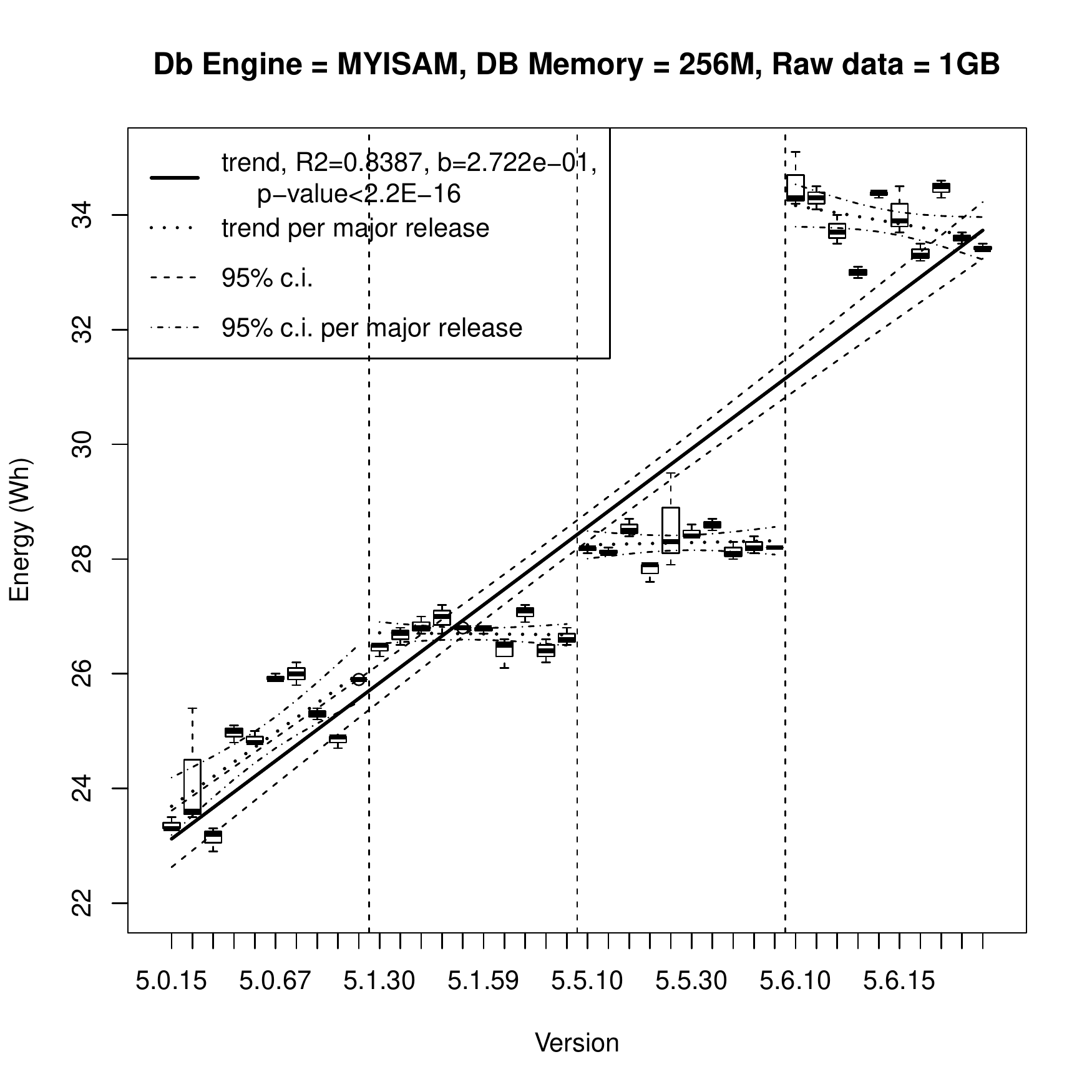}
    \caption{Experiment 1}
    \label{fig:exp1-bp}
  \end{subfigure}
  \begin{subfigure}[b]{0.5\textwidth}
    \includegraphics[width=\textwidth]{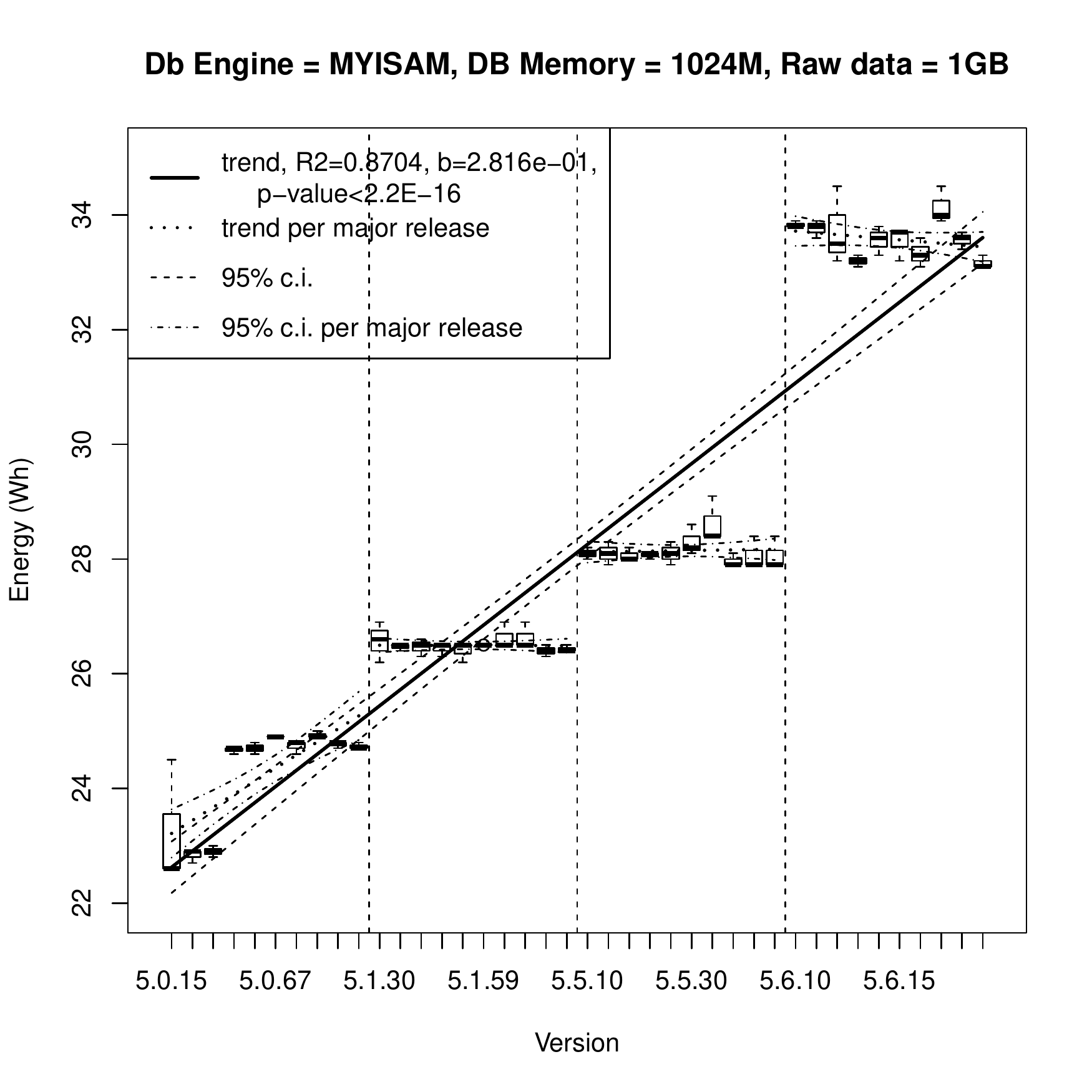}
    \caption{Experiment 3}
    \label{fig:exp3-bp}
  \end{subfigure}
  \begin{subfigure}[b]{0.5\textwidth}
    \includegraphics[width=\textwidth]{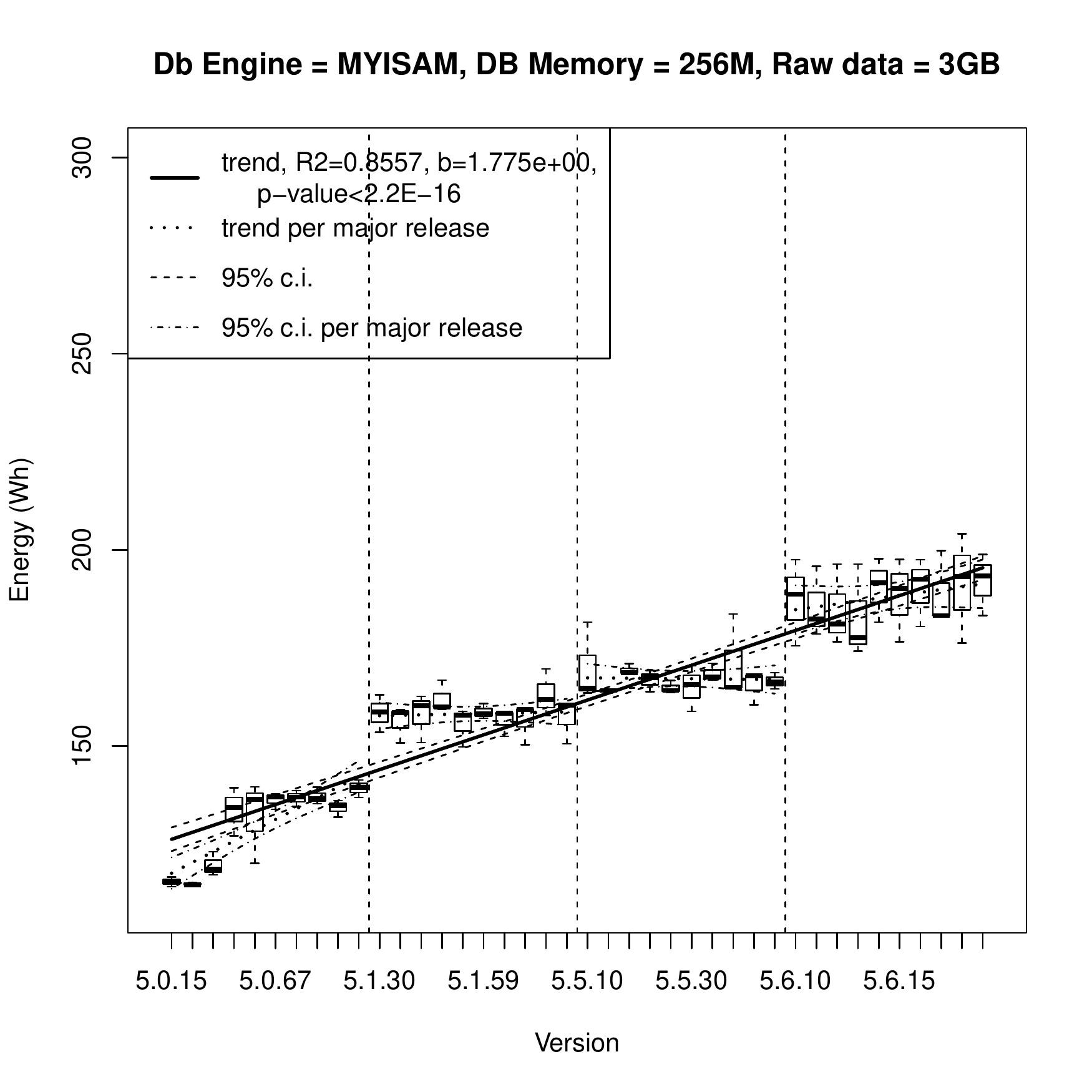}
    \caption{Experiment 2}
    \label{fig:exp2-bp}
  \end{subfigure}
  \begin{subfigure}[b]{0.5\textwidth}
    \includegraphics[width=\textwidth]{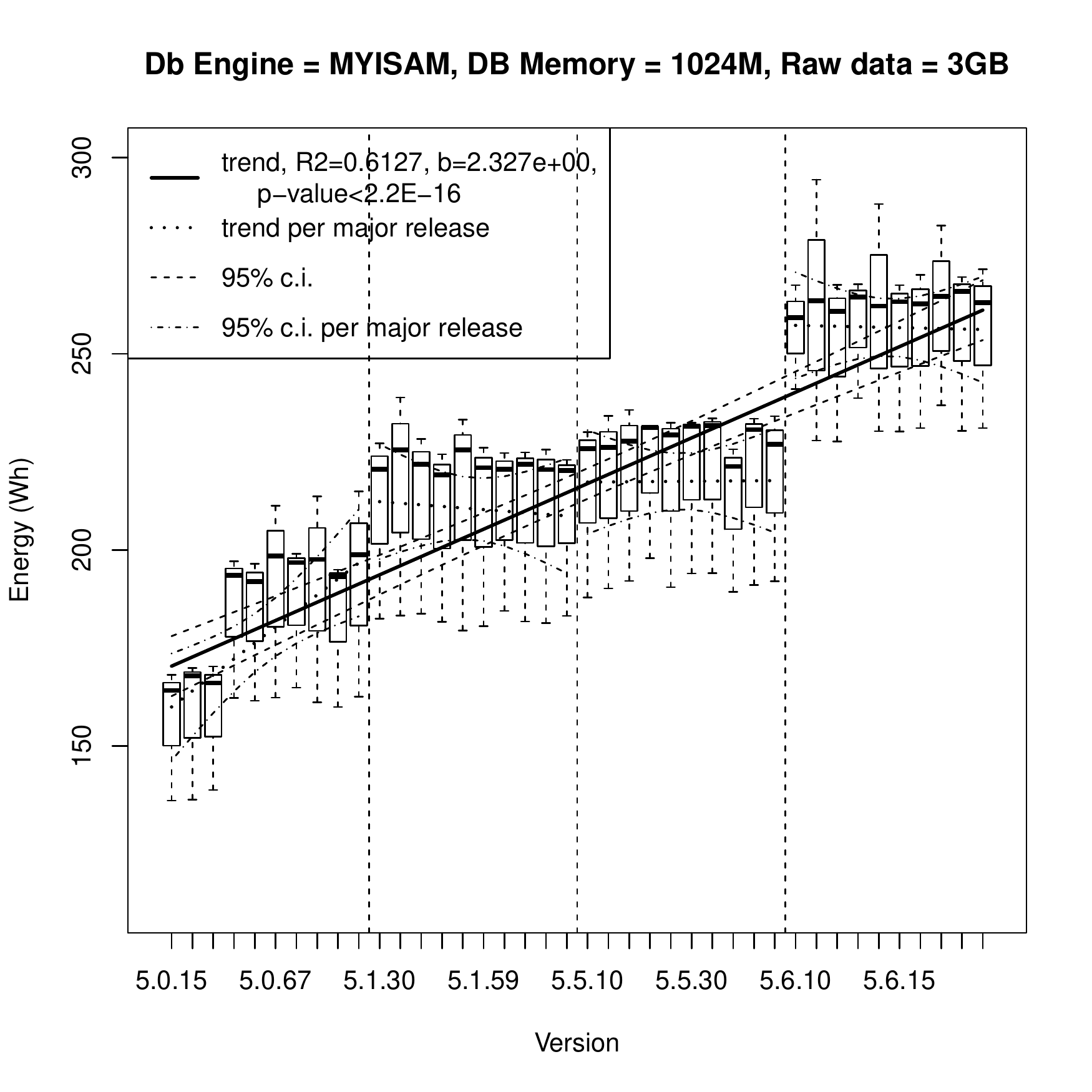}
    \caption{Experiment 4}
    \label{fig:exp4-bp}
  \end{subfigure}

    \caption{Experiments: box plot of the energy consumed in MySQL MyISAM's versions. In subplots' solid line depicts a trend line obtained using linear regression; the dashed lines show the 95\% confidence interval of the trend line. The dotted lines represent a trend line (obtained using linear regression) per major release, and the dot-dashed lines show the 95\% confidence interval of these trend lines per major release. Vertical long-dash lines show boundaries between major releases.}
    \label{fig:energy_vs_version_exp_1-4}
\end{figure}

\begin{figure}[ht]
  \begin{subfigure}[b]{0.5\textwidth}
    \includegraphics[width=\textwidth]{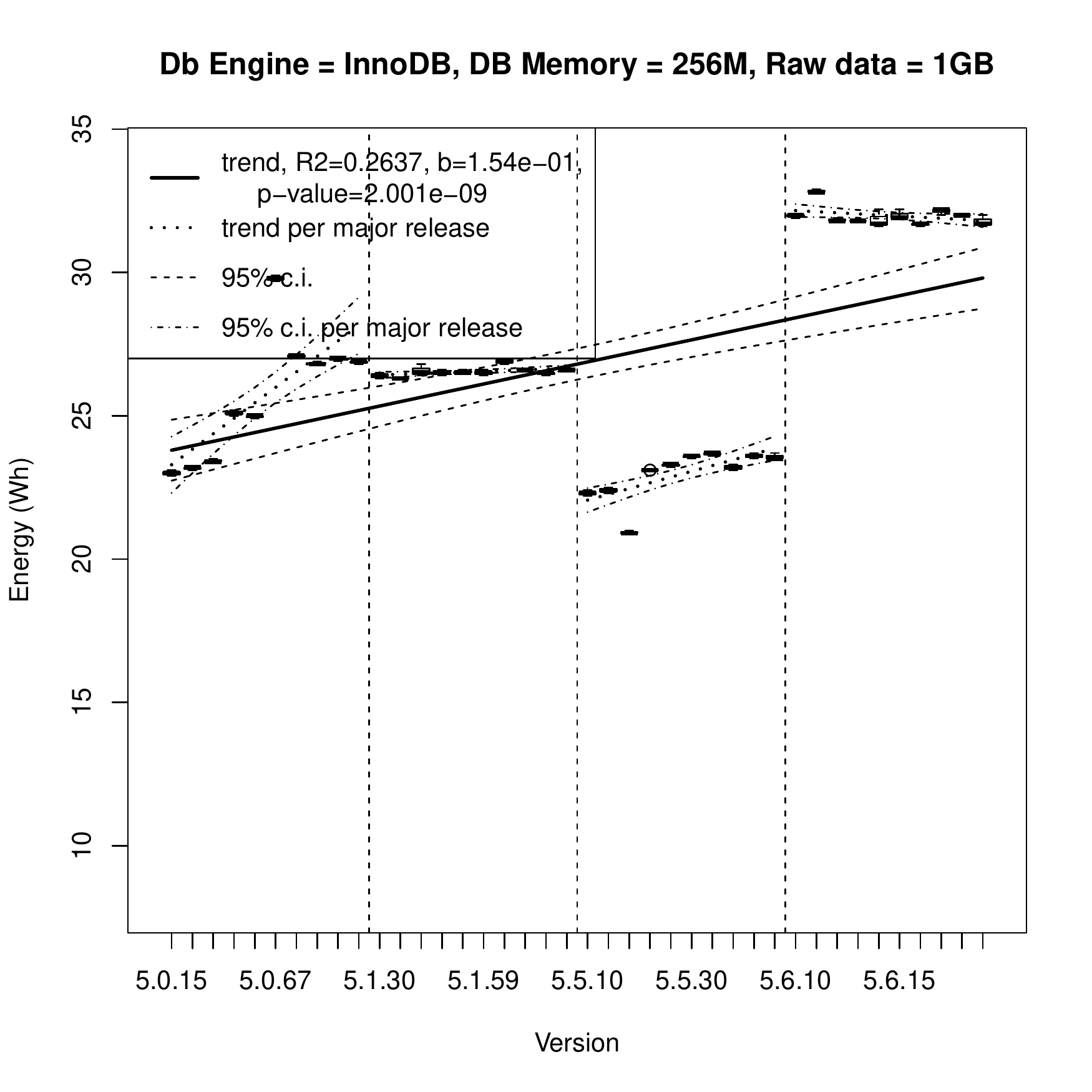}
    \caption{Experiment 5}
    \label{fig:exp5-bp}
  \end{subfigure}
  \begin{subfigure}[b]{0.5\textwidth}
    \includegraphics[width=\textwidth]{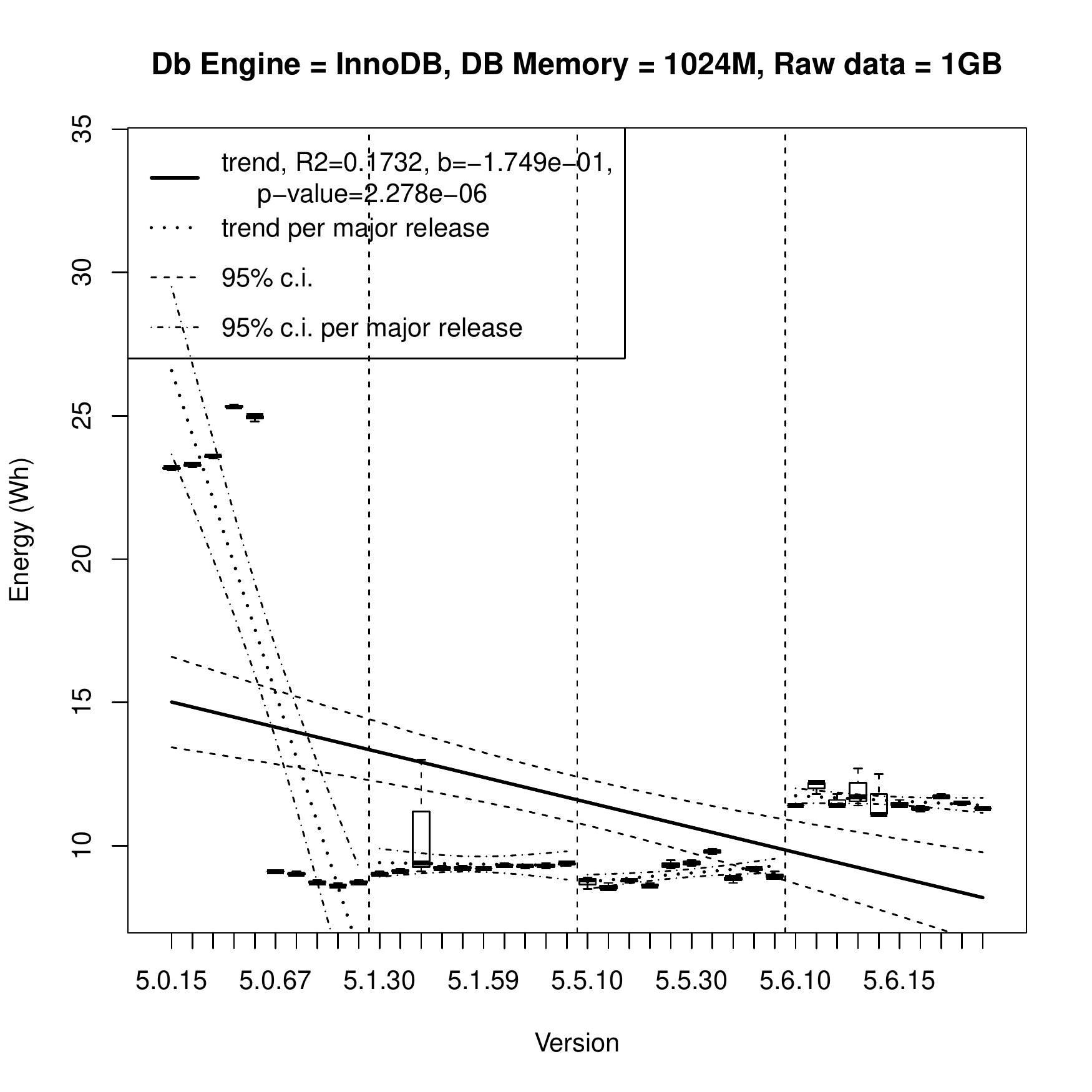}
    \caption{Experiment 7}
    \label{fig:exp7-bp}
  \end{subfigure}
  \begin{subfigure}[b]{0.5\textwidth}
    \includegraphics[width=\textwidth]{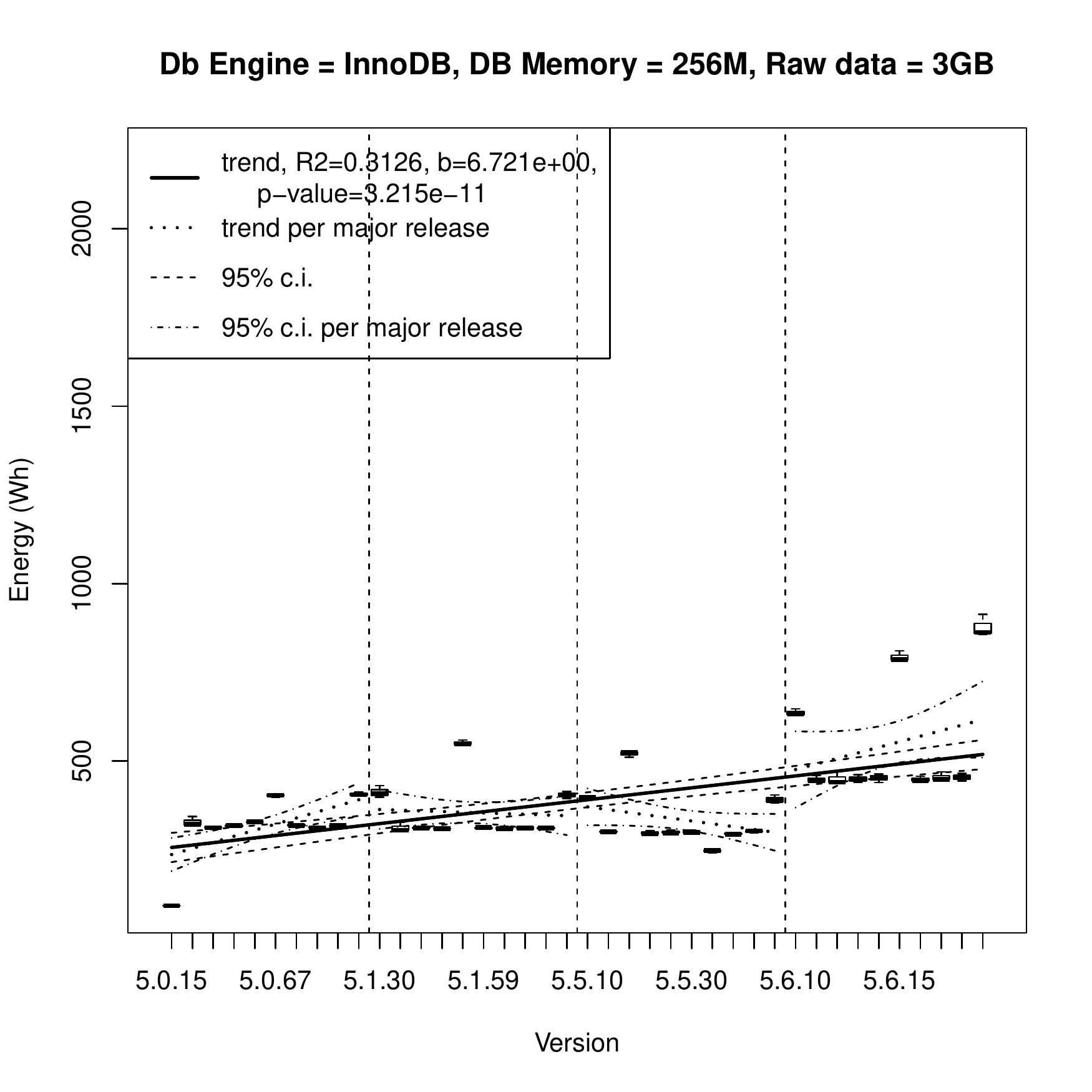}
    \caption{Experiment 6}
    \label{fig:exp6-bp}
  \end{subfigure}
  \begin{subfigure}[b]{0.5\textwidth}
    \includegraphics[width=\textwidth]{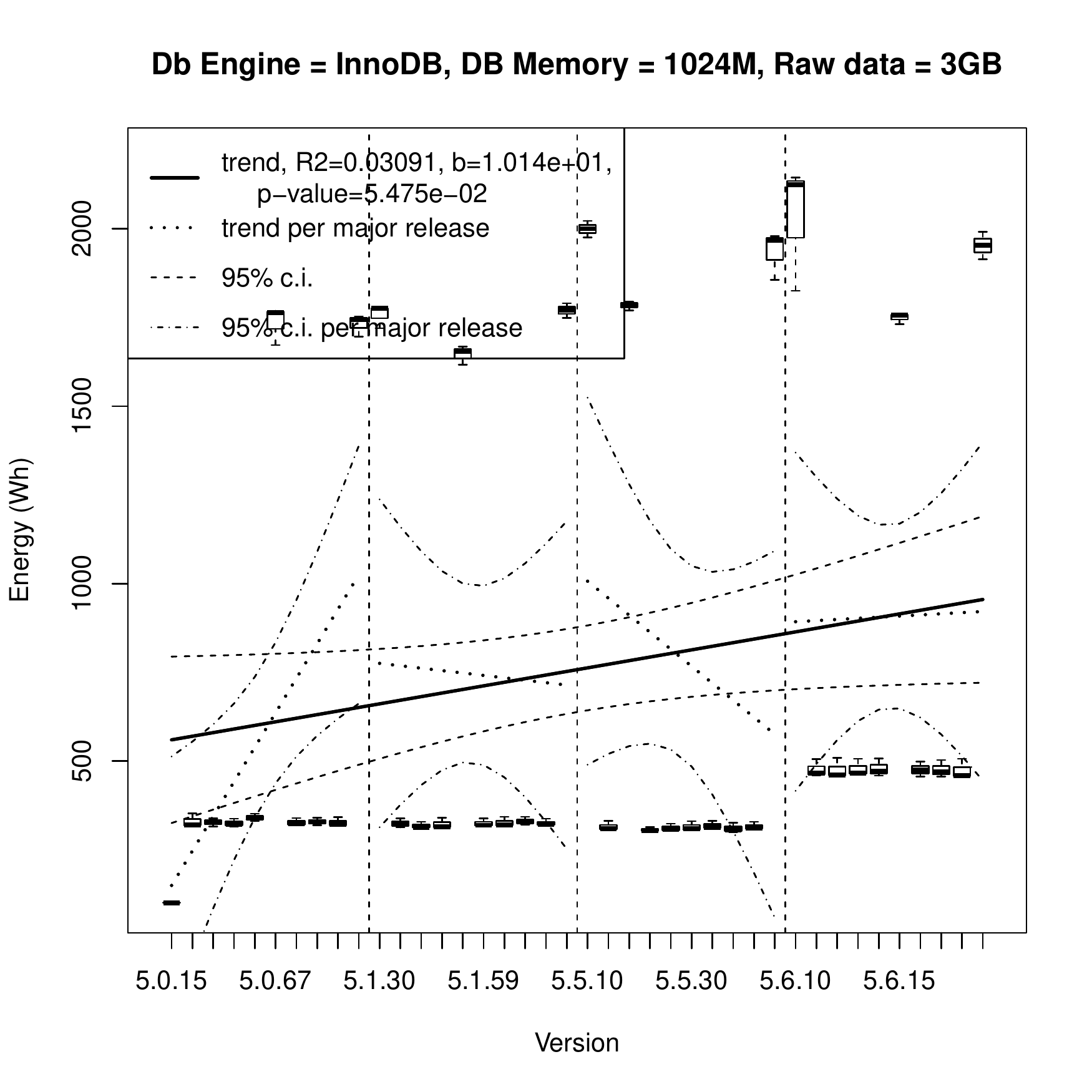}
    \caption{Experiment 8}
    \label{fig:exp8-bp}
  \end{subfigure}

    \caption{Experiments: box plot of the energy consumed in MySQL InnoDB's versions. In subplots' solid line depicts a trend line obtained using linear regression; the dashed lines show the 95\% confidence interval of the trend line. The dotted lines represent a trend line (obtained using linear regression) per major release, and the dot-dashed lines show the 95\% confidence interval of these trend lines per major release. Vertical long-dash lines show boundaries between major releases.}
    \label{fig:energy_vs_version_exp_5-8}
\end{figure}

\begin{figure}[ht]
  \begin{subfigure}[b]{0.5\textwidth}
    \includegraphics[width=\textwidth]{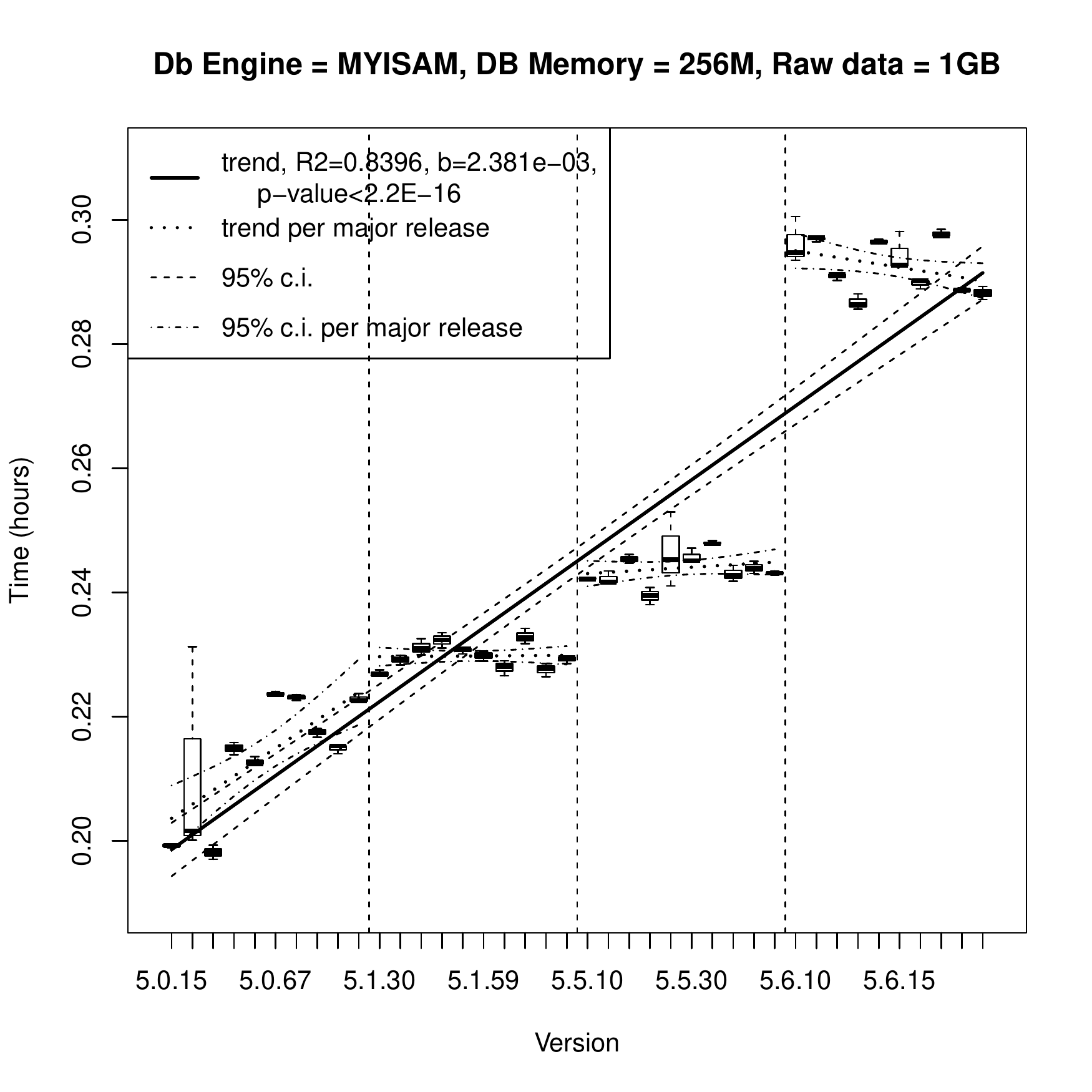}
    \caption{Experiment 1}
    \label{fig:exp1-et}
  \end{subfigure}
  \begin{subfigure}[b]{0.5\textwidth}
    \includegraphics[width=\textwidth]{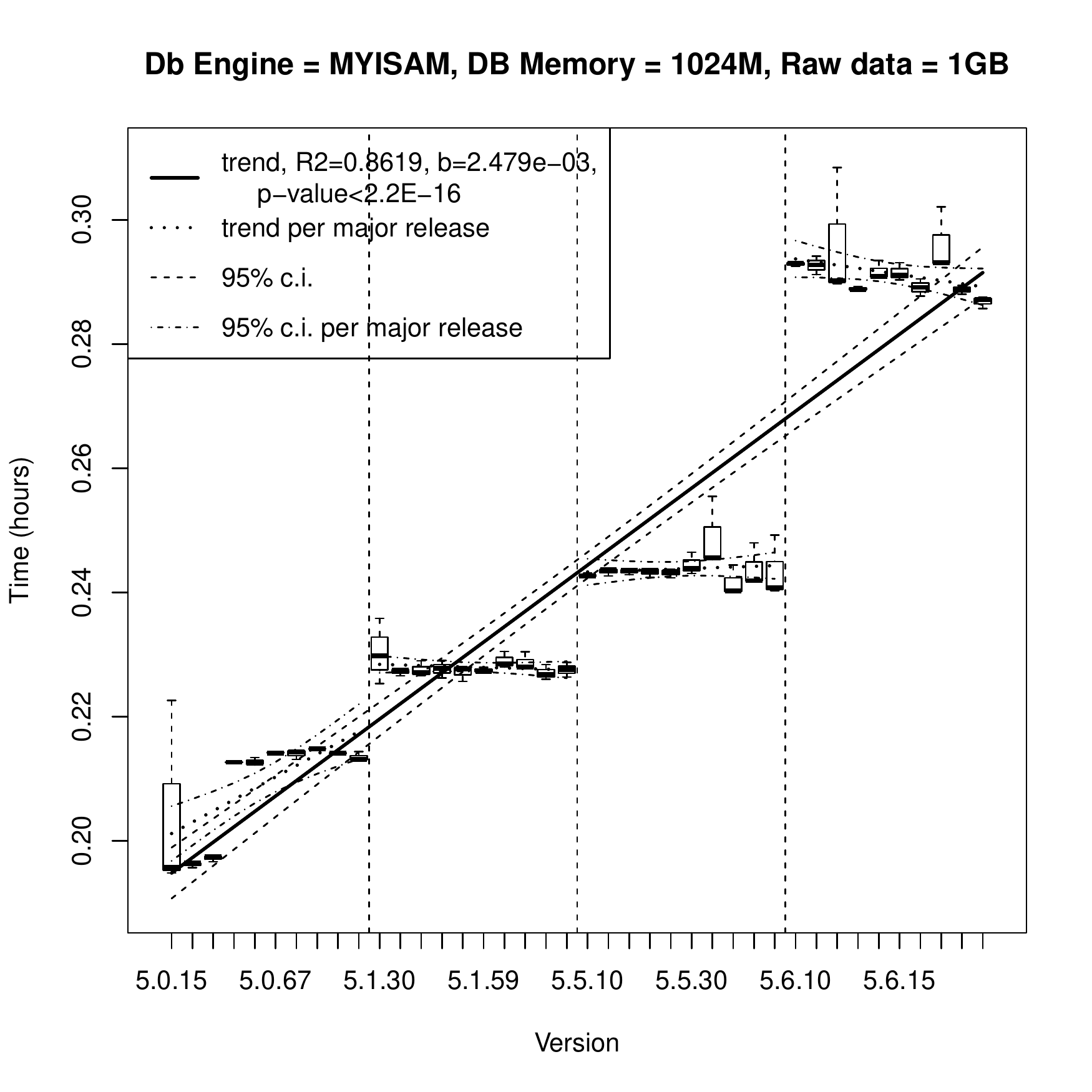}
    \caption{Experiment 3}
    \label{fig:exp3-et}
  \end{subfigure}
  \begin{subfigure}[b]{0.5\textwidth}
    \includegraphics[width=\textwidth]{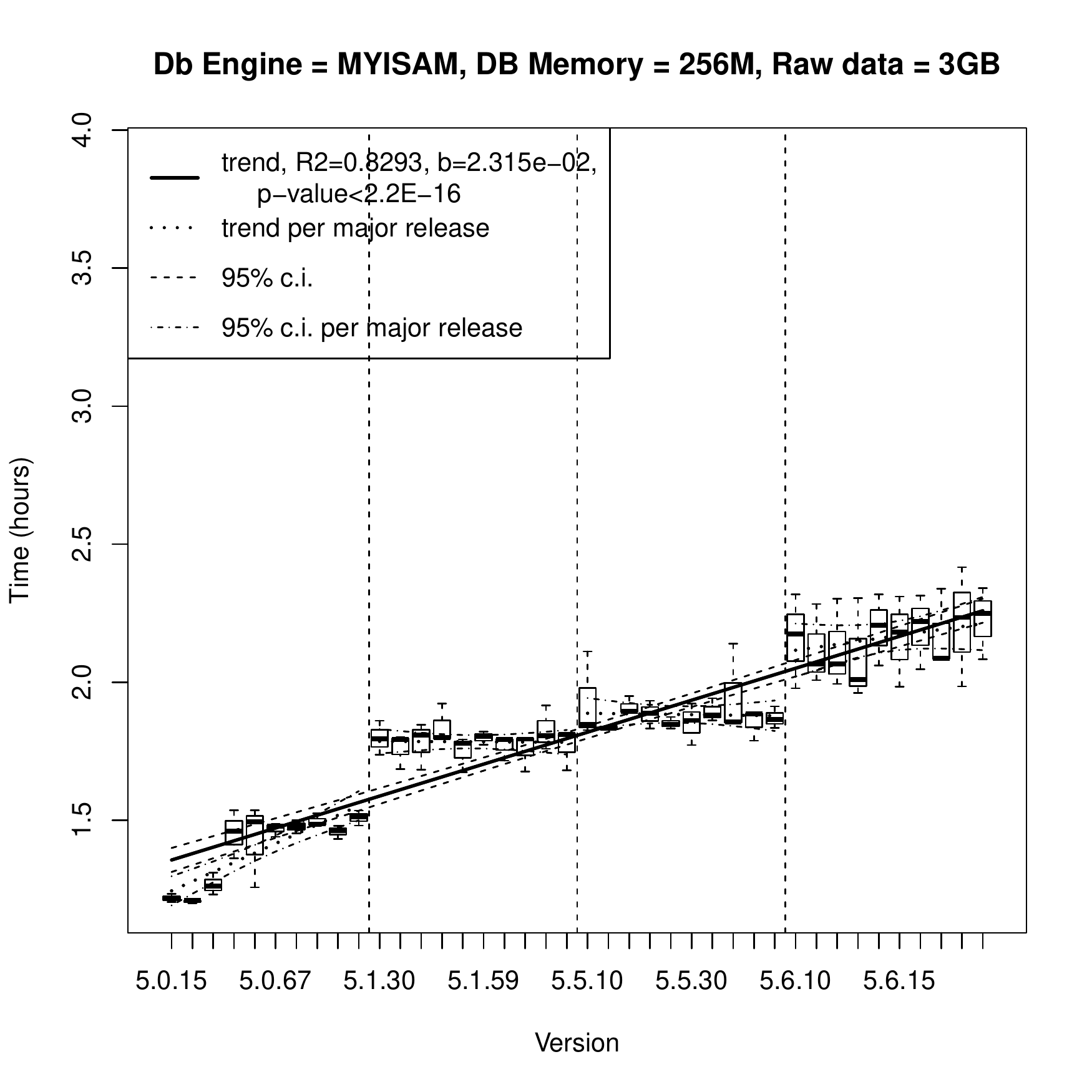}
    \caption{Experiment 2}
    \label{fig:exp2-et}
  \end{subfigure}
  \begin{subfigure}[b]{0.5\textwidth}
    \includegraphics[width=\textwidth]{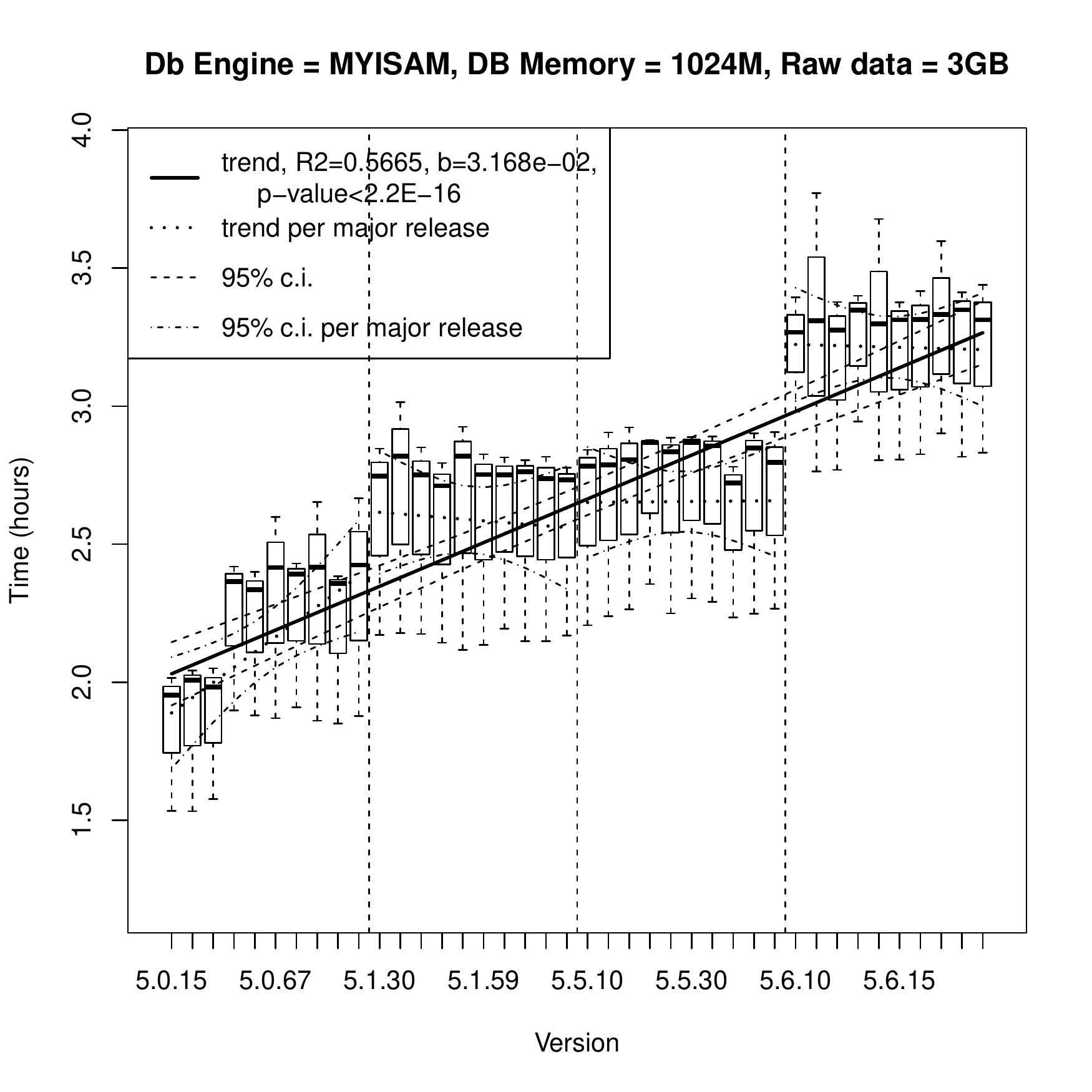}
    \caption{Experiment 4}
    \label{fig:exp4-et}
  \end{subfigure}

    \caption{Experiments: box plot of the execution time in MySQL MyISAM's versions. In subplots' solid line depicts a trend line obtained using linear regression; the dashed lines show the 95\% confidence interval of the trend line. The dotted lines represent a trend line (obtained using linear regression) per major release, and the dot-dashed lines show the 95\% confidence interval of these trend lines per major release. Vertical long-dash lines show boundaries between major releases.}
    \label{fig:time_vs_version_exp_1-4}
\end{figure}

\begin{figure}[ht]
  \begin{subfigure}[b]{0.5\textwidth}
    \includegraphics[width=\textwidth]{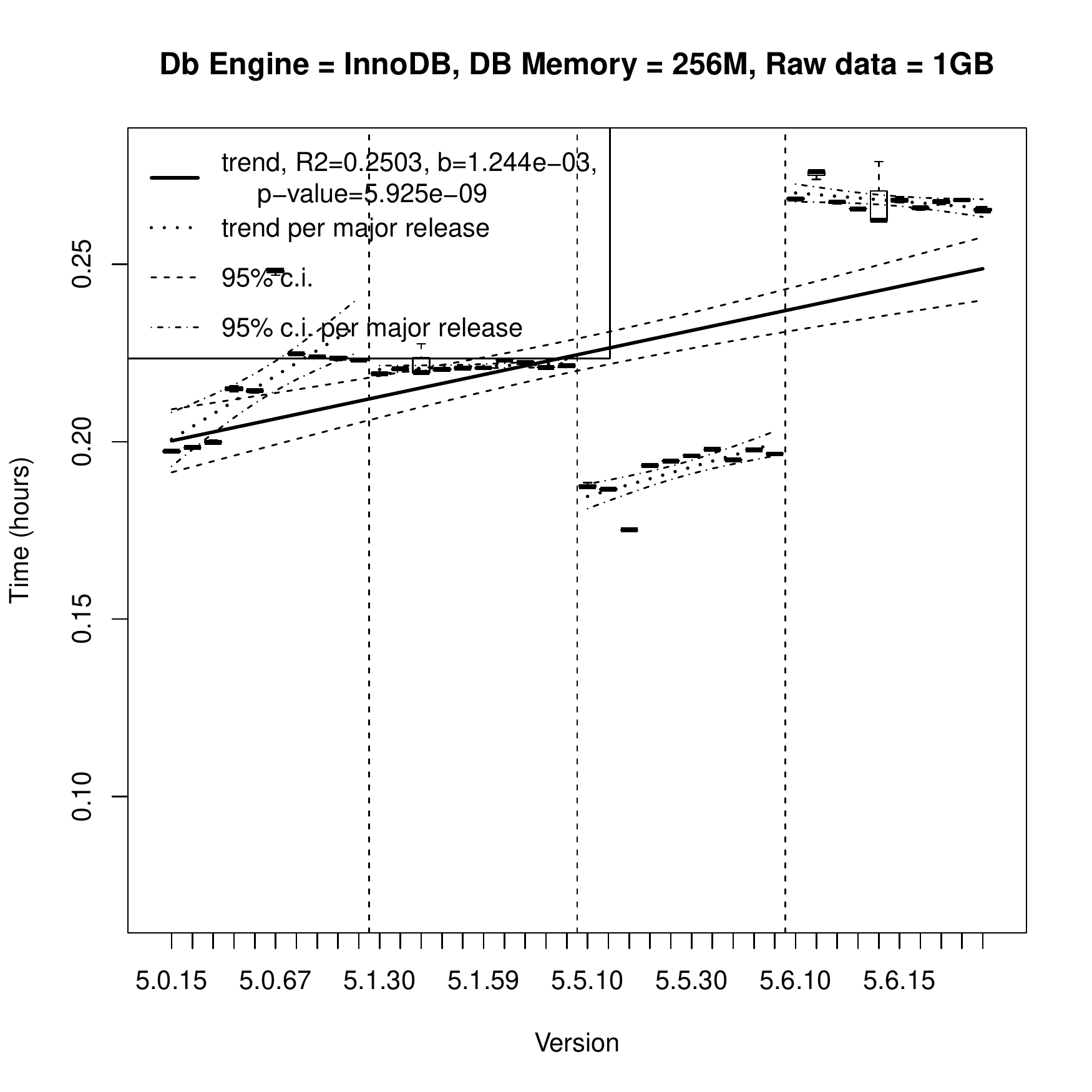}
    \caption{Experiment 5}
    \label{fig:exp5-et}
  \end{subfigure}
  \begin{subfigure}[b]{0.5\textwidth}
    \includegraphics[width=\textwidth]{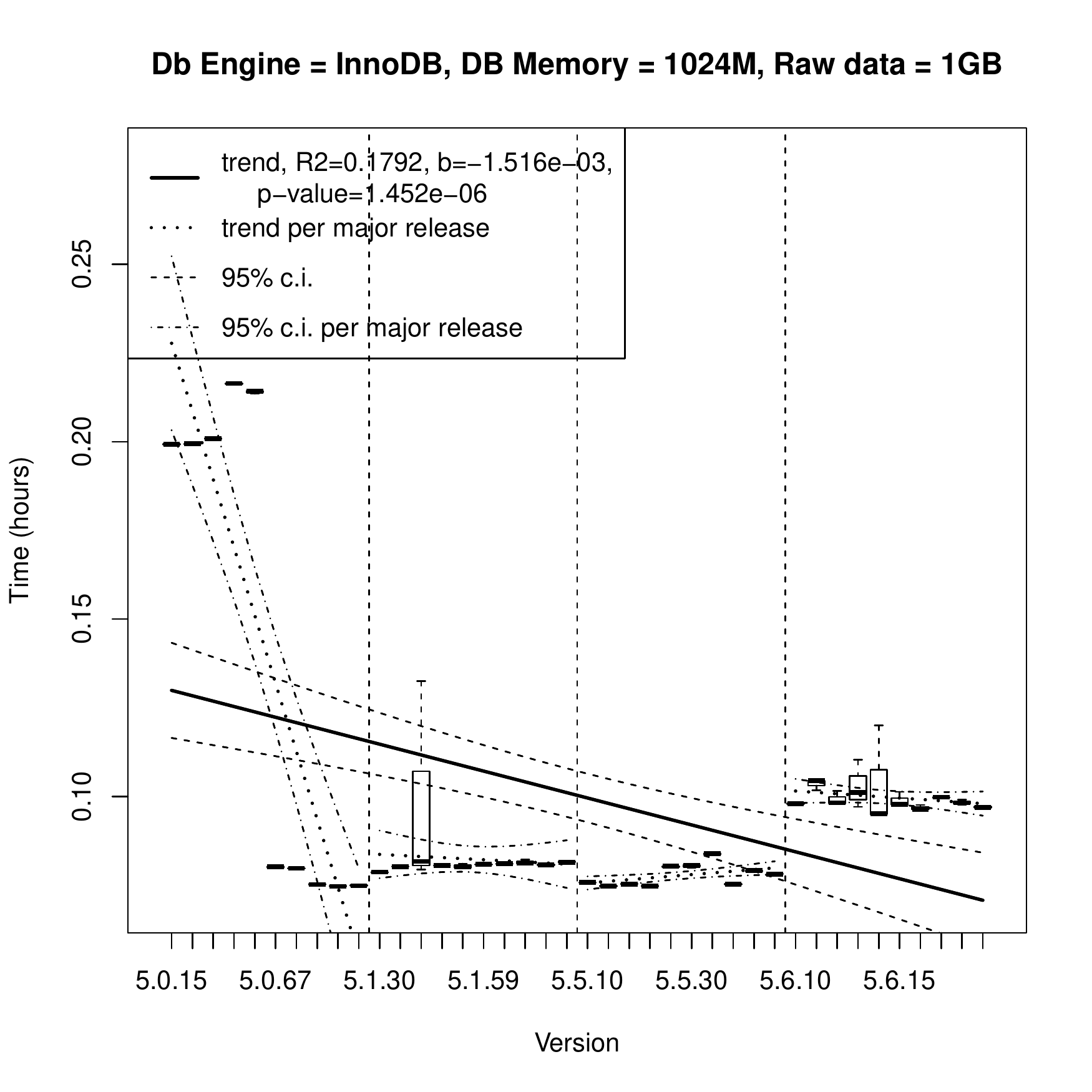}
    \caption{Experiment 7}
    \label{fig:exp7-et}
  \end{subfigure}
  \begin{subfigure}[b]{0.5\textwidth}
    \includegraphics[width=\textwidth]{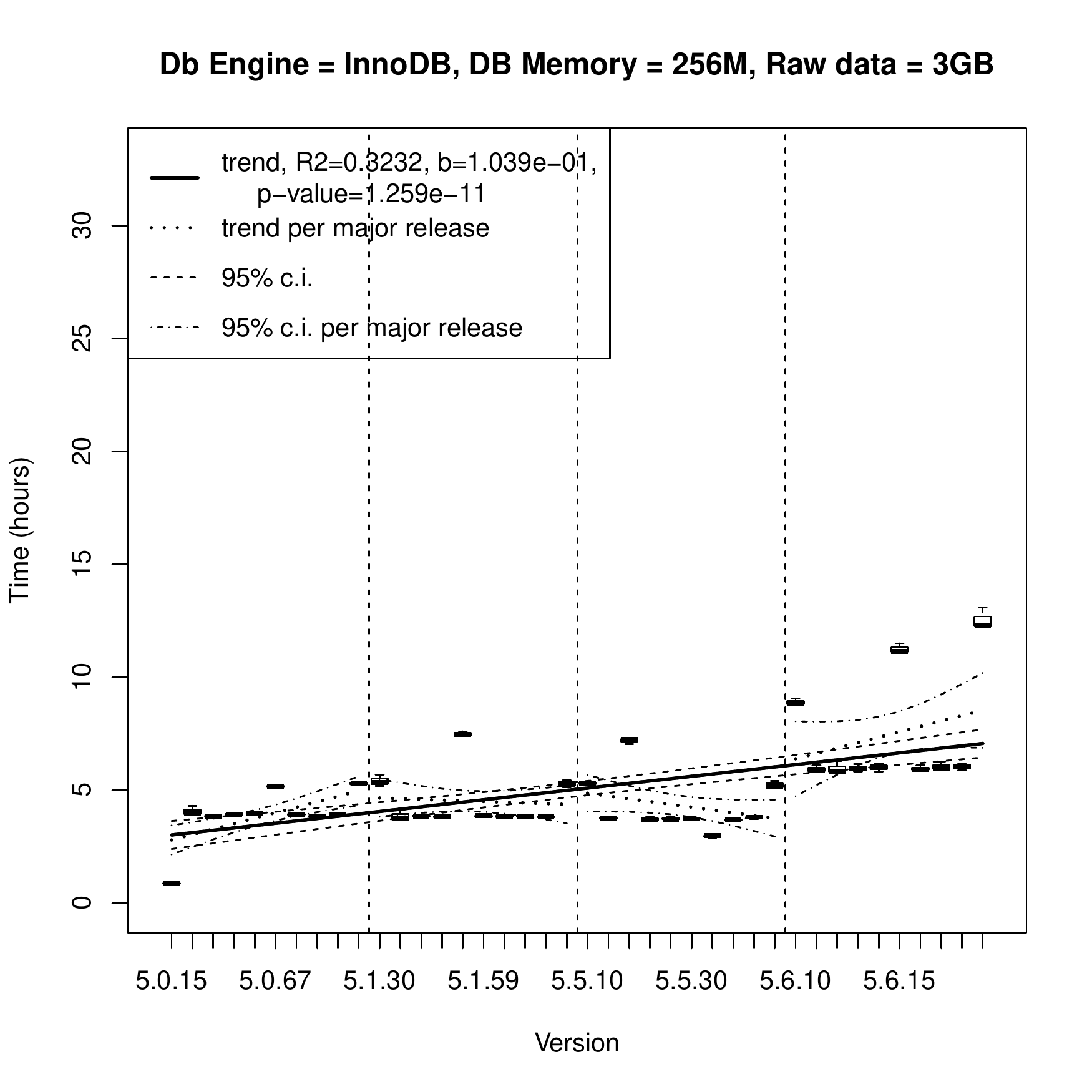}
    \caption{Experiment 6}
    \label{fig:exp6-et}
  \end{subfigure}
  \begin{subfigure}[b]{0.5\textwidth}
    \includegraphics[width=\textwidth]{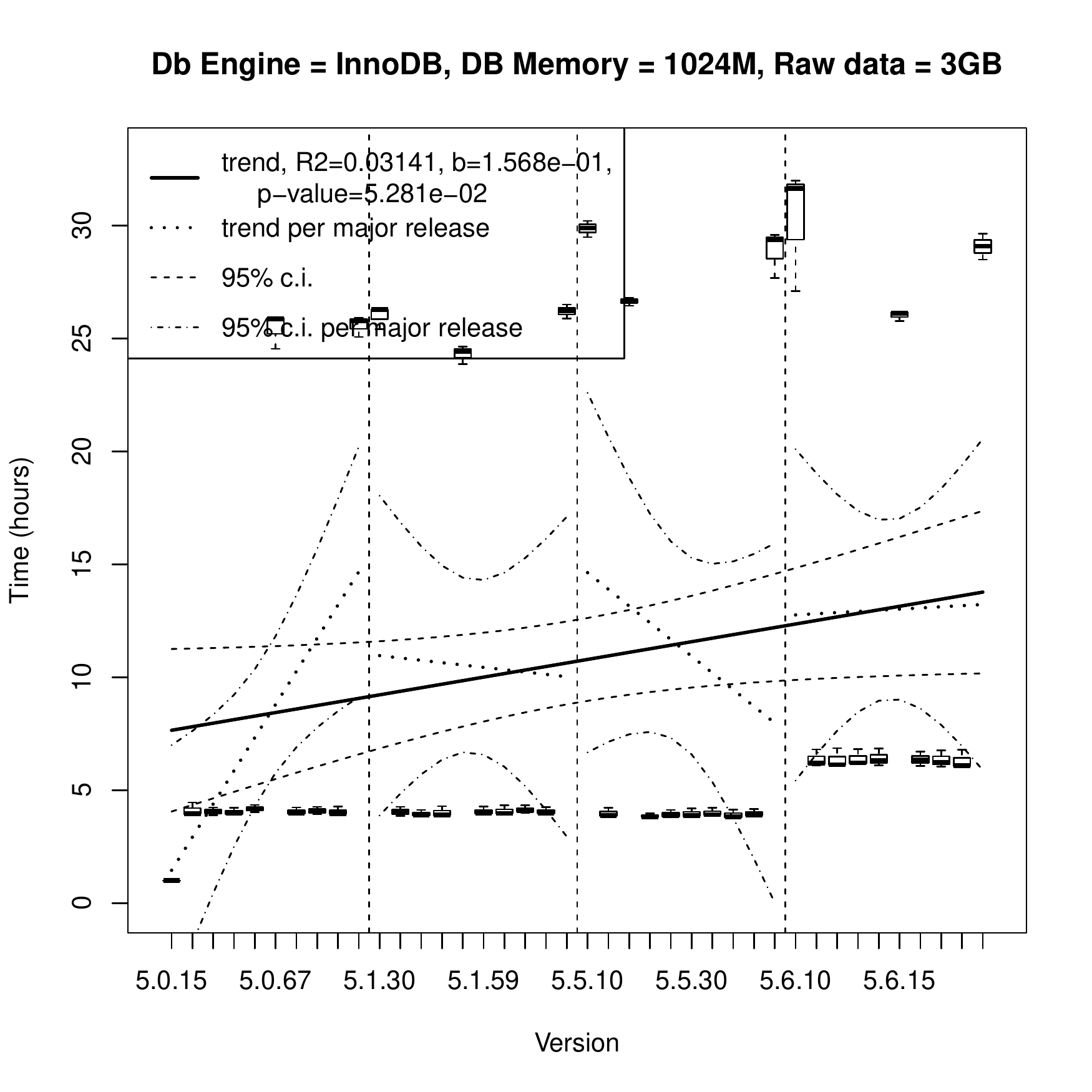}
    \caption{Experiment 8}
    \label{fig:exp8-et}
  \end{subfigure}

    \caption{Experiments: box plot of the energy consumed in MySQL InnoDB's versions. In subplots' solid line depicts a trend line obtained using linear regression; the dashed lines show the 95\% confidence interval of the trend line. The dotted lines represent a trend line (obtained using linear regression) per major release, and the dot-dashed lines show the 95\% confidence interval of these trend lines per major release. Vertical long-dash lines show boundaries between major releases.}
    \label{fig:time_vs_version_exp_5-8}
\end{figure}

This appendix contains Figures used in Section~\ref{sec:results}. Figures~\ref{fig:energy_vs_version_exp_1-4} and \ref{fig:time_vs_version_exp_1-4} show energy consumed and time spent by experiments executed against MySQL MyISAM engine. Figures~\ref{fig:energy_vs_version_exp_5-8} and \ref{fig:time_vs_version_exp_5-8} -- against MySQL InnoDB engine.

\bibliography{ref}
\bibliographystyle{wileyj}

\end{document}